\documentclass{article}

\usepackage[preprint]{neurips_2025}

\usepackage[utf8]{inputenc} 
\usepackage[T1]{fontenc}    
\usepackage{hyperref}       
\usepackage{url}            
\usepackage{booktabs}       
\usepackage{amsfonts}       
\usepackage{nicefrac}       
\usepackage{microtype}      
\usepackage{xcolor}         

\usepackage{bm}             
\usepackage{graphicx}
\usepackage{subcaption}
\usepackage{amsmath}

\usepackage{pgfplots}
\pgfplotsset{compat=1.18}
\usepackage{pgfplotstable}

\title{Convergence of physics-informed neural networks modeling time-harmonic wave fields}

\author{%
  Stefan Schoder\\
  Institute of Fundamentals and Theory of Electrical Engineering\\
  Graz University of Technology\\
  8010 Graz, Austria \\
  \texttt{stefan.schoder@tugraz.at} \\
  \And
  Aneta Furmanová and Viktor Hruška \\
  Faculty of Electrical Engineering \\
  Czech Technical University in Prague\\
  Technická 1902/2 Praha, Czech Republic\\
  \texttt{furmaane@fel.cvut.cz}, 
  \texttt{viktor.hruska@fel.cvut.cz} \\
}

\begin{document}

\maketitle

\begin{abstract}
  Studying physics-informed neural networks (PINNs) for modeling partial differential equations to solve the acoustic wave field has produced promising results for simple geometries in two-dimensional domains. One option is to compute the time-harmonic wave field using the Helmholtz equation. Compared to existing numerical models, the physics-informed neural networks forward problem has to overcome several topics related to the convergence of the optimization toward the "true" solution. The topics reach from considering the physical dimensionality (from 2D to 3D), the modeling of realistic sources (from a self-similar source to a realistic confined point source), the modeling of sound-hard (Neumann) boundary conditions, and the modeling of the full wave field by considering the complex solution quantities. Within this contribution, we study 3D room acoustic cases at low frequency, varying the source definition and the number of boundary condition sets and using a complex speed of sound model to account for some degree of absorption. We assess the convergence behavior by looking at the loss landscape of the PINN architecture, the $L^2$ error compared to a finite element reference simulation for each network architecture and configuration. The convergence studies showed that at least six training points per wavelength are necessary for accurate training and subsequent predictions of the PINN. The developments are part of an initiative aiming to model the low-frequency behavior of room acoustics, including absorbers.
\end{abstract}

\section{Introduction}
Modelling of physical fields involving wave-related phenomena represents common ground for many disciplines spanning from acoustics \citep{pierce2019acoustics}, fluid dynamics \citep{bailly2015turbulence} and various branches of mechanics \citep{Kaltenbacher2015Numerical}, over gravitational waves \citep{maggiore2008gravitational} or electrodynamics \citep{griffiths2023introduction} to applications in thermodynamics \citep{sieniutycz2002variational} and reaction kinetics \citep{ortoleva1974variety}. In quantum mechanics \citep{schrodinger2003collected}, the wave function is important to describe the probability distribution of particles.

Mathematically, the time-harmonic counterpart of the (d’Alembertian) linear wave equation is the Helmholtz equation ($\Delta p + k^2 p = g$), which we will consider here for the case of longitudinal waves. While many aspects pertain to the fundamental nature of wave solutions and are transferable across domains, our focus here is specifically on acoustic pressure waves. In acoustics, the Helmholtz equation forms the foundation for analyzing a wide range of scenarios, particularly in applications such as room acoustics (see e.g. \citep{Kraxberger2023Finite}).

The reader can find many details on the design of PINNs for solving the Helmholtz equation in a very recent article by Veerababu and Ghosh \cite{Veerababu2025}, who look at solving the Helmholtz equation using PINNs in two dimensions.
Compared to existing literature on PINNs for solving the Helmholtz equation, we study the convergence of PINN training. The literature on this topic does not address the selected room acoustic case of the Helmholtz equation (3D and Neumann boundary conditions --  see e.g., \cite{YeonjongShin2020, YulingJiao2022, Doumche2025, Jahaninasab2024, DeRyck2022, Pantidis2023}). Shin et al. \cite{YeonjongShin2020} and Jiao et al. \cite{YulingJiao2022} made remarks on general cases of elliptical PDEs, but the closer treatment and computational examples cannot be directly related to room acoustics.

In the following, we consider a unit cube $\Omega = [0,1]^3$ containing a medium possessing a complex wave speed of the form $c^2 = c_0^2 + i c_\mathrm{i}^2$ with $1/k^2 = c^2/\omega^2 = 1/k_0^2 + i/k_\mathrm{i}^2$, with $\omega = 2\pi f$. A ratio of the imaginary to the real part of the complex wave speed is defined $\eta = c_\mathrm{i}^2/c_0^2 = -0.04$. We made use of the example from the library DeepXDE \citep{lu2021deepxde} and adapted it to the 3D room acoustics example \citep{schoder2025physics}. The resulting inhomogeneous Helmholtz equation of the acoustic pressure $p = p_\mathrm{r} + ip_\mathrm{i} \in \mathbb{C}$ can be expressed by one equation for the real part of $p_\mathrm{r} \in \mathbb{R}$ and the imaginary part $p_\mathrm{i} \in \mathbb{R}$
\begin{align}
    \frac{\Delta}{k_0^2}p_\mathrm{r} - \frac{\Delta}{k_\mathrm{i}^2}p_\mathrm{i} + p_\mathrm{r} &= - g_\mathrm{r} + \eta  g_\mathrm{i} \quad &&\text{on } \bm x \in \Omega \\
    \frac{\Delta}{k_0^2}p_\mathrm{i} + \frac{\Delta}{k_\mathrm{i}^2}p_\mathrm{r} + p_\mathrm{i} &= -\eta g_\mathrm{r} - g_\mathrm{i} \quad &&\text{on } \bm x \in \Omega
\end{align}
with Neumann boundary conditions 
\begin{align}
    \nabla p_\mathrm{r} \cdot \bm{n} = 0 \quad \mathrm{on}\quad &&\bm x \in \partial\Omega\\
    \nabla p_\mathrm{i} \cdot \bm{n} = 0 \quad \mathrm{on}\quad &&\bm x \in \partial\Omega
\end{align}
on the six faces of the cube, with the outward normal $\bm n$ on the boundary $\partial\Omega$ of the domain $\Omega$. The separate analysis of the real and the imaginary part allows us to leverage the standard data processing routines of a GPU using real numbers instead of the complex number. The real and imaginary part of the forcing is denoted as $g_\mathrm{r}$ and $g_\mathrm{i}$, respectively.

We exploit the fact that we can choose the forcing so that the equation has an analytical solution. Specifically, for the forcing $g_\mathrm{r} = 2\cos(kx)\cos(ky)\cos(kz) = 2p_0$ and $g_\mathrm{i} = 0$ the solution is for $\eta\ll 1$ approximately $p_\mathrm{r} = (1-\frac{3\eta^2}{4})p_0 \approx p_0$ and $p_\mathrm{r} = -\frac{\eta}{2 +9\eta^2}p_0 \approx -\frac{\eta}{2}p_0 $. Regarding this equation, the aim of this study is to provide more insight over the convergence properties of neural networks for modelling the wave solution based on PINNs as a forward problem. Therefore, we defined four objectives:
\begin{enumerate}
    \item Assess the convergence behavior for PINN architectures for the two-dimensional and three-dimensional problem using the loss landscape, number of iterations for convergence onset, and final converge regarding specific network hyperparameters.
    \item Study on how much training points per wavelength are necessary to obtain accurate results of the wave field for a specific wave field frequency.
    \item For more illustrative comparison to previous literature \citep{schoder2025physics}, the convergence behavior is studied for varying source sharpness parameters $\hat{s}$, yielding a source function of
    \begin{align} \label{eq:forcing}
        g_\mathrm{r} = 2\cos(kx)\cos(ky)\cos(kz)\mathrm{e}^{-\frac{||\bm{x}  - \bm{x}_s||_2^2}{2s^2}} \;,
    \end{align}
    using $\bm{x} = (x,y,z)$ and a source location $\bm{x}_s$.
    \item Finally, we study whether Transfer learning from a Green's function solution helps the PINN converge during training.
\end{enumerate}

\section{Methods}

\subsection{Outline of the PINN architecture}

The architecture employed in this work is the feed-forward neural network (FNN) with $L$ layers and $N_{\ell}$ neurons in the $\ell$-th layer.
The network is trained to approximate the mapping $\bm{x} \rightarrow p(\bm{x})$ and
the loss function to be minimized is given as $\mathcal{L} = w_{\mathrm{PDE}} \mathcal{L}_{\mathrm{PDE}} +
    w_{\mathrm{BC}} \mathcal{L}_{\mathrm{BC}}$,
where $ w_{\mathrm{PDE}},  w_{\mathrm{BC}}$ stand for the weights of the loss terms linked with the underlying partial differential equation and its boundary conditions, respectively.
The loss terms are
\begin{align}
    \label{eq:loss_pde}
    \mathcal{L}_{\mathrm{PDE}} &= \frac{1}{N_{\mathrm{PDE}}} \sum^{N_{\mathrm{PDE}}} || r_{\mathrm{PDE}} (\bm{x}) ||^2 ~,
    \quad \forall \bm{x} \in \Omega ~,
    \\
    \label{eq:loss_pde_bc}
    \mathcal{L}_{\mathrm{BC}} &=  \frac{1}{N_{\mathrm{BC}}} \sum^{N_{\mathrm{BC}}} || r_{\mathrm{BC}} (\bm{x}) ||^2 ~,
    \qquad ~~ \forall \bm{x} \in \partial \Omega ~,
\end{align}
and $r_{\mathrm{PDE}}, r_{\mathrm{BC}} $ being the respective residuals. 
The default chosen activation function is $\sin$ as suggested in \cite{schoder2025physics}, unless specified otherwise.
The definition of data points, the optimizer and other hyperparameters are specified in the appendix \ref{sec:TA:2dsetup}.
For this study, we employed the open-source library DeepXDE \citep{lu2021deepxde} with PyTorch backend. The computer architecture and code version specification are provided in the technical appendix \ref{sec:tech_appendix}.

\subsection{Convergence assessment and loss landscape}
A common issue of PINNs is that the non-trivial loss function is ill-conditioned, which results in optimizer struggling with convergence \citep{rathore2024}.
Several training strategies have been introduced to address the ill-conditioning,
such as 
incorporating sparse data as a regulator \citep{Gopakumar2023},
locally adaptive activation functions with slope recovery \citep{jagtap2020locally},
residual-based adaptive refinement \citep{lu2021deepxde},
transfer learning \citep{LIU2023112291, Chen2021, GOSWAMI2020102447} or warm-up training \citep{GarciaWarmUp}.
Nonetheless, according to \cite{rathore2024}, no single strategy is known to enhance the PINN performance of any arbitrary PDE.

To assess the impact of proposed modifications on convergence, it is insufficient to compare the time taken for the model to converge. A more insightful approach involves analyzing the convexity of the loss function by analyzing the eigenvalues of the loss Hessian \citep{loss_landscape}. For this purpose, the code from \cite{Gopakumar2023} was employed here.

\subsection{Proposed physics informed strategies to improve convergence}

In this work, we study the PINNs convergence and intend to improve it by integrating knowledge about the underlying physics. The improvements concern mainly the 3D case, which suffers from slower convergence and higher computational demand as opposed to 2D case (cf. \cite{Veerababu2025}). 
We focus on two elements of the PINN training: the features of the activation function and a possibility of pretraining on data obtained from the use of the free-field Green's function.

\subsubsection{Scaled activation functions} \label{sec:scaled}

Jagtap \textit{et al.} \cite{jagtap2020locally} suggest increasing the learning capacity of the network by employing locally adaptive activation functions (LAAF) with slope recovery.
The main idea is that the additional adaptive parameters alter the loss landscape dynamically, which should improve the convergence mainly at the early training stages. On the other hand, the slope recovery term further complicates the loss term (and consequently poses an additional threat to the convexity of the loss function). Moreover, the LAAF has not yet been implemented in the employed library DeepXDE for the chosen PyTorch backend. However, we realized that the problem could be more fundamental.

Here we would like to point out an important physical aspect: functions that are solutions of the wave equations (at least in their strong formulation) shall be twice differentiable. Moreover, the property that the second derivatives of functions are generally non-zero is one of the basic features of the description of wave motion. Hence, in the analogous machine learning problem, we need to be able to calculate the second derivative of the neural network that result in a nonzero value. This reasoning naturally excludes a whole class of activation functions of being considered in the first place. For instance, the popular ReLU function is piecewise linear, so its second derivative is zero everywhere, which disables the learning \citep{Maczuga2023}.

With our chosen activation function $\sin(x)$, it is possible that the training is encountering the same issue: if the inputs of the activation functions are too small, the sine would be operating in a linear regime and, as a consequence, the zero second derivative would prevent the network from learning. Note that this is likely to happen, because the spatial coordinates (i.e. input of the PINN) are from (0,1) and the Glorot initialization (see the Appendix) prevent the network weights of reaching high initial values. Therefore, solely changing the wavenumber-like parameter of the activation function ($\sin(kx)$ with $k > 1$) might improve the overall convexity and accelerate the convergence onset without needing additional adaptive parameters to learn.

\subsubsection{Discrepancy learning with Green's function} \label{sec:discrepancy}

The prior knowledge of underlying physics can be incorporated in more ways than just through the loss function. Rather than training the model from scratch, one can pretrain it as a simple feed-forward neural network on related data.
We can leverage the free-space Green's functions (GF), commonly used to solve inhomogeneous PDEs, and allow the PINN to learn just the discrepancy arising from the fact that the GF assumes an unbounded domain, while our problem is defined on a finite domain with specific boundary conditions.
This approach is called discrepancy learning \citep{ebers2023, kaheman2019}.

The free space Green's function for inhomogeneous Helmholtz equation in 3D 
and an impulsive point source $\delta$ at $\bm{x}_0$ reads
\citep{delfs}
\begin{align}
    G(\bm{x}, \bm{x}_0) = \frac{\exp{(i k ||\bm{x}  - \bm{x}_0||_2)}}{4 \pi ||\bm{x}  - \bm{x}_0||_2} ~.
\end{align}

The solution $p_\mathrm{GF}$ for an inhomogeneous Helmholtz equation with source term of strength $g$ (see Eq. \eqref{eq:forcing}) is then
\begin{align} \label{eq:pgf}
    p_\mathrm{GF}(\bm{x}) = \int_{-\infty}^{\infty} G(\bm{x}, \bm{x}_0) g(\bm{x}, \bm{x}_0) \mathrm{d}V ~.
\end{align}

The overall training then looks as follows. First, a classical NN of the same architecture is pretrained in a supervised manner on data generated by the solution $p_\mathrm{GF}$ from Eq. \eqref{eq:pgf}.
The integral is implemented as a sum over the whole domain $\Omega$.
The loss function in this case is just the mean squared error between the predicted pressure and the true pressure (i.e. without any involvement of either the differential equation, nor its boundary conditions).
Then, we switch to training the PINN in an unsupervised manner (with the loss function defined in Eqs. \eqref{eq:loss_pde} and \eqref{eq:loss_pde_bc}),
while some of the layers might be frozen.

\subsection{Comparison to FEM and Green's function reference solution}

The PINN results are compared to a reference finite element simulation in openCFS 24.03 \citep{Schoder2022openCFS}, and the data processing for comparison is done in pyCFS \citep{wurzinger2024pycfs}. We assess convergence by the relative $ L^2$ error
\begin{equation}
    e_\mathrm{rel,FEM} = \sqrt{\frac{
                         \sum_{j=1}^{|\mathcal{S}_{\mathrm{error}}|} \left( p_{j,\mathrm{FEM}} - p_{j,\mathrm{PINN}} \right)\left( p_{j,\mathrm{FEM}} - p_{j,\mathrm{PINN}} \right)^*}
                        {\sum_{j=1}^{|\mathcal{S}_{\mathrm{error}}|} \left( p_{j,\mathrm{FEM}} \right)\left( p_{j,\mathrm{FEM}} \right)^*} }  \, ,
                        \label{eq:error}
\end{equation}
where $|\mathcal{S}_{\mathrm{error}}|$ is the number of $ L^2$ error evaluation points inside the domain $\Omega$ and $()^*$ is the complex conjugate. A PINN has learned a meaningful solution when
\begin{equation}
    e_\mathrm{rel,FEM} < e_\mathrm{rel,GF} = \sqrt{\frac{
                         \sum_{j=1}^{|\mathcal{S}_{\mathrm{error}}|} \left( p_{j,\mathrm{GF}} - p_{j,\mathrm{PINN}} \right)\left( p_{j,\mathrm{GF}} - p_{j,\mathrm{PINN}} \right)^*}
                        {\sum_{j=1}^{|\mathcal{S}_{\mathrm{error}}|} \left( p_{j,\mathrm{GF}} \right)\left( p_{j,\mathrm{GF}} \right)^*} } \, ,
                        \label{eq:cond}
\end{equation}
is less than the error exhibited by convoluting the free-fields Green's function with the source function $g_\mathrm{r}$ \citep{rucz2024analysis}.

\section{Results}

The results for the 2D case $\Omega = [0,1]^2$m$^2$ and $\hat{s}\rightarrow \infty$ (setup details are given in the appendix \ref{sec:TA:2dsetup}), show for a real value of the speed of sound $e_\mathrm{rel,FEM} = \{0.10, 0.64, 1.74\} \%$ and for the complex speed of sound $e_\mathrm{rel,FEM} = \{0.026, 0.18, 0.42\} \%$ (convergence onset at 4000 iterations). The total training time was about 7.5 minutes and 12 minutes, respectively. When splitting the four faces of the boundary into separate Neumann boundary condition sets (eight compared to two for the complex speed of sound case), the number of iterations when the PINN started to converge was slightly later (at around 6\,000 iterations compared to 4\,000 iterations), however the obtained relative error was in an acceptable range $e_\mathrm{rel,FEM} = \{0.45, 0.59, 1.71\} \%$. The final error values are obtained after 70\,000 iterations (training time of about 15 minutes). The respective loss landscapes can be found in the appendix figure \ref{fig:lossL2d}. All three optimization problems show a regular convex loss landscape.

\subsection{3D cube case with forcing $\hat{s}\rightarrow \infty$}

The results for the 3D case $\Omega = [0,1]^3$m$^3$ and $\hat{s}\rightarrow \infty$ (setup details are given in the appendix \ref{sec:TA:3dsetup}), exhibit the relative errors provided in table \ref{tab:T1} after 200\,000 iterations using the Adam optimizer. It is particularly interesting that for the base setup (wavelength of about $0.5$m), with increasing scaling of the activation function argument, the PINN starts to converge after fewer iterations. Similar observations were made by \cite{jagtap2020locally} for the locally adaptive activation function formulation. The relative error $e_\mathrm{rel,FEM}$ was comparably low for all the predictions carried out and in a similar range as for the two dimensional case. When using 6 boundary condition sets compared to one, the network parameter optimization onset happened later in the optimization. However, comparable relative error was achieved at the end of the 200\,000 iterations.

\begin{table}[h!]
\centering
\caption{Summary of convergence onset iterations and $e_\mathrm{rel,FEM}$ for different complex 3D cube case with forcing $\hat{s}\rightarrow \infty$. The PINN (200\,000 iterations), with one boundary condition set (1BC) and six boundary condition sets (6BC), with sinus activation functions scaled. The standard supervised learned networks (50\,000 iterations) based on the solution datasets are denoted by (NN), $t$ denotes the training time.}\label{tab:T1}
\begin{tabular}{lrrrcccc}
\toprule
\textbf{Experiment Name} & \multicolumn{3}{c}{\textbf{Onset}} & \multicolumn{3}{c}{\textbf{$e_\mathrm{rel,FEM}$ (\%)}} & $t$ (min) \\
\toprule
PINN, 1BC, $\sin(x)$      & 12000  & 25000  & 16000   & 0.00034 & 0.039 & 0.047 & 63\\
PINN, 6BC, $\sin(x)$      & 35000  & 29000  & 29000  & 0.051   & 0.136 & 0.059 & 71\\
PINN, 6BC, $\sin(2x)$     & 18000  & 19000  & 15000  & 0.058 & 0.074 & 0.043 &74\\
PINN, 6BC, $\sin(4x)$      & 3000   & 4000  & 2000  & 0.039 & 0.053 & 0.027 &75\\
PINN, 6BC, $\sin(8x)$      & 2000   & 2000  & 2000  & 0.074 & 0.078 & 0.20 &75\\
NN, $\sin(x)$   & 9000  & 8000  & 36000  & 0.020 & 0.051 & 3.60 & 1:13\\
NN, $\sin(2x)$  & 7000  & 19000  & 9000  & 0.028 & 0.033 & 0.10 &3:25\\
NN, $\sin(4x)$  & 1000  & 1000  & 1000  & 0.013 & 0.052 & 0.97 &3:25\\
NN, $\sin(8x)$  & 1000  & 1000  & 1000  & 9.13 & 9.72 & 13.00 &3:25\\ 
\bottomrule
\end{tabular}
\end{table}

As a benchmark for the PINN training, we also trained the same network architecture in a supervised manner on the analytic result data (NN key in table \ref{tab:T1}). The data was obtained by the analytical solution of the $\hat{s}\rightarrow \infty$ case (see derivation in the introduction). As expected, the equivalent PINN models converged slower than the purely data-trained network. Overall, the same level of convergence of the purely data-trained networks was reached after fewer total iterations (the error values at 50\,000 final iterations were presented for the data models).

\subsubsection{How many training points per wavelength are needed}
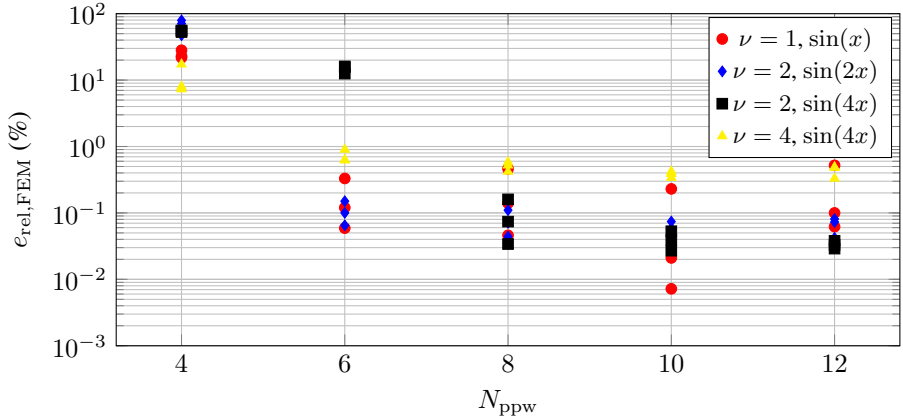
\begin{figure}[h!]
\centering
\begin{tikzpicture}
\begin{axis}[
    width=12cm,
    height=6cm,
    xlabel={$N_\mathrm{ppw}$},
    ylabel={$e_\mathrm{rel,FEM}$ (\%)},
    legend style={at={(0.99,0.99)}, anchor=north east, font=\small},
    grid=both,
    ymode=log,
    log basis y={10},
    xtick={4,6,8,10,12},
    ymin=0.001, ymax=100,
    cycle list name=color list,
    only marks,
    mark options={solid},
]

\addplot+[mark=*] coordinates {(4,21.79) (6,0.059) (8,0.046) (10,0.0072) (12,0.062) (4,28.16) (6,0.12) (8,0.46) (10,0.23) (12,0.52) (4,22.97) (6,0.33) (8,0.14) (10,0.021) (12,0.10)};
\addlegendentry{$\nu = 1$, $\sin(x)$}

\addplot+[mark=diamond*, only marks] coordinates {
    (4, 47.48) (6, 0.065) (8, 0.044) (10, 0.043) (12, 0.042)
    (4, 70.84) (6, 0.10) (8, 0.071) (10, 0.058) (12, 0.073)
    (4, 79.30) (6, 0.15) (8, 0.11) (10, 0.074) (12, 0.081)
};
\addlegendentry{$\nu = 2$, $\sin(2x)$}

\addplot+[mark=square*] coordinates {(4,52.90) (6,12.51) (8,0.034) (10,0.027) (12,0.029) (4,55.47) (6,16.06) (8,0.16) (10,0.039) (12,0.038) (4,55.72) (6,15.60) (8,0.074) (10,0.053) (12,0.033)};
\addlegendentry{$\nu = 2$, $\sin(4x)$}

\addplot+[mark=triangle*, solid] coordinates {(4,7.31) (6,0.62) (8,0.42) (10,0.34) (12,0.33) (4,8.00) (6,0.63) (8,0.52) (10,0.39) (12,0.49) (4,17.29) (6,0.90) (8,0.58) (10,0.42) (12,0.48)};
\addlegendentry{$\nu = 4$, $\sin(4x)$}

\end{axis}
\end{tikzpicture}
\caption{Relative errors ($e_\mathrm{rel,FEM}$) from the PINN as a function of the points per wavelength and for variable frequencies $\nu$ using three recorded error values from tables \ref{tab:PINN_ppw_3d_inf}.}\label{fig:ppw_s_infty}
\end{figure}

In this section, we report the outcome of the study on how many training points are needed to obtain converged results of the PINN predictions. In particular, we varied the dimensionless frequency $\nu = f L/c_0$ (where $L$ is the length of the cube) in a computationally feasible set $\nu = \{1, 2,4 \}$. These frequencies are typical values for wave-based simulations in room acoustics \citep{Kraxberger2023Validated}
. For each frequency, the Helmholtz equation is solved by using randomly sampled training points to evaluate the loss function. The total number $N_\mathrm{Total}$ of the training points is based on the wave number and called points per wavelength $N_\mathrm{ppw}$
\begin{equation}
     N_\mathrm{Total} = (N_\mathrm{ppw} c_0/f)^3\, .
\end{equation}
Figure~\ref{fig:ppw_s_infty} shows the dependency of the relative error in dependence on the points per wavelength. For each parameter combination, three optimization runs were executed and the errors reported. For $\nu = 1$, the sinus activation function was used and the tendency shows that at around six points per wavelength the PINN starts to converge to a sufficiently small relative error. In the case of $\nu = 2$ and a $\sin(2x)$ activation function, a similar convergence trend as for $\nu = 1$ is recovered. In the case of $\nu = 2$ and a $\sin(4x)$ activation function, at least eight points per wavelength are needed to obtain an error below 1\%. A similar picture was obtained for $\nu = 4$ and a $\sin(4x)$ activation function. For numerical methods and in certain cases, the error is starting to stay in an acceptable range if one uses six points per wavelength \cite{marburg2002six} or more. The obtained results (presented in figure~\ref{fig:ppw_s_infty}) are comparable to this classical rule of thumb, so one may not expect a PINN to converge if the resolution is less.

\begin{figure}[h!]
\centering
\begin{tikzpicture}
\begin{axis}[
    width=12cm,
    height=5.5cm,
    xlabel={$N_\mathrm{ppw}$},
    ylabel={$e_\mathrm{rel,FEM}$ (\%)},
    legend style={at={(0.99,0.99)}, anchor=north east, font=\small},
    grid=both,
    ymode=log,
    log basis y={10},
    xtick={4,6,8,10,12},
    ymin=0.01, ymax=10,
    cycle list name=color list,
    only marks,
    mark options={solid},
]

\addplot+[only marks, mark=*] coordinates {(4,1.75) (6,0.12) (8,0.065) (10,0.023) (12,0.023) (4,0.59) (6,0.094) (8,0.056) (10,0.057) (12,0.10)(4,2.35) (6,0.11) (8,0.082) (10,0.071) (12,0.055)};
\addlegendentry{No layers frozen}

\addplot+[only marks, mark=square*] coordinates {(4,1.85) (6,0.85) (8,1.17) (10,0.73) (12,0.59) (4,1.23) (6,0.39) (8,0.50) (10,0.39) (12,0.45) (4,9.70) (6,4.17) (8,4.09) (10,4.16) (12,8.52)};
\addlegendentry{Frozen, except the first 2}

\addplot+[only marks, mark=triangle*] coordinates {(4,0.79) (6,0.18) (8,0.090) (10,0.062) (12,0.062) (4,0.50) (6,0.11) (8,0.099) (10,0.090) (12,0.090) (4,2.84) (6,0.085) (8,0.12) (10,0.12) (12,0.11)};
\addlegendentry{Frozen, except the last 2}

\end{axis}
\end{tikzpicture}
\caption{Relative errors ($e_\mathrm{rel,FEM}$) from the discrepancy learning as a function of the points per wavelength and frequency $\nu=2$ using three recorded error values from tables \ref{tab:PINN_discrepancy_ppw_3d_inf}.}\label{fig:ppw_s_infty_DL}
\end{figure}
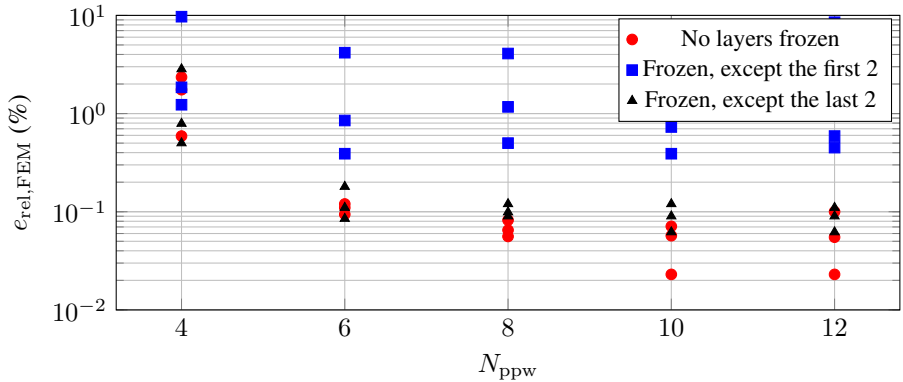

From now on, we focus on the frequency $\nu = 2$. When doing discrepancy learning, first the fully connected neural network is pre-trained with a free-field acoustic solution based on a fast computation using the free-field Green's function. In figure~\ref{fig:ppw_s_infty_DL}, the follow-up training of the pre-trained network with the PDE and BC loss term is presented after 40\,000 iterations. Compared to the ordinary PINN training, this procedure takes substantially less training time. The investigation shows that training from an educated initial pre-trained network state is beneficial. A minor difference in the relative error was observed between the case in which all the trainable parameters were free to adapt and the case in which only the last two layers were trained. We can extract from figure~\ref{fig:ppw_s_infty_DL} that the relative error is already strongly reduced for discrepancy learning at four points per wavelength (within a range of one percent) compared to more than 10\% in figure~\ref{fig:ppw_s_infty}.

\subsection{3D cube case with forcing $\hat{s}=1$}

In the next study, we look at the details when setting the source shape parameter to $\hat{s}=1$. In that case, the solution function cannot be determined easily analytically. Therefore, we used a finite element reference solution with the same example case definitions. The relative error (at 200\,000 iterations) is evaluated by considering the result of the PINN and the finite element simulation in the node locations of the finite element reference. The neural network named "V" has a shape where the number of neurons increases in the layers from the input to the output. The version V $\sin(x)$ only uses the $\sin(x)$ activation function. The three hidden layers have 32, 64, and 128 neurons. For versions (Va, Vb): $\sin(x)$, $\sin(2x)$, and $\sin(4x)$ activation function in the hidden layers respectively. The version Va uses a $\tanh(x)$ activation function and the Vb a $\sin(x)$ for the output layer. The shape was motivated by the experience that the layers closer to the input learn the global characteristics and the layers closer to the output the more fine-grained structure \cite{markidis2021old}. Compared to the standard PINN setup with three layers of 150 neurons each and variable activation functions $\sin(x)$, $\sin(2x)$, $\sin(4x)$, and $\sin(8x)$, the Va structure led to comparable relative errors and slightly less training time. The errors of the investigated network setups are reported in table \ref{tab:PINN_s1}. All V networks performed similarly. During the training point resolution study, we found that even at four points per wavelength the relative error was below 2.5\%, and with increasing resolution it dropped further.

\begin{table}[ht]
\centering
\caption{Summary of convergence error $e_\mathrm{rel,FEM}$ and training time $t$ for different complex 3D cube case with forcing $\hat{s}=1$ after 200\,000 iterations. }
\label{tab:PINN_s1}
\begin{tabular}{lccccc}
\toprule
\textbf{Name} & \textbf{$N_\mathrm{ppw}$} & \multicolumn{3}{c}{\textbf{$e_\mathrm{rel,FEM}$ (\%)}} & $t$ (s)\\
\midrule
PINN, $\sin(x)$                 & 10     & 0.1260 & 0.0347 & 0.0457 & 4240 \\ 
PINN, $\sin(2x)$                & 10     & 0.1514 & 0.1634 & 0.0390 & 4459\\ 
PINN, $\sin(4x)$                & 10     & 0.0244 & 0.5545 & 0.0507 & 4523\\ 
PINN, $\sin(8x)$                & 10     & 0.1134 & 0.0990 & 0.0633 & 4527\\ 
PINN, Va                        & 10     & 0.0785 & 0.0971 & 0.0559 & 3508 \\ 
PINN, Vb                        & 10     & 0.1905 & 0.1353 & 0.1066 & 3464\\ 
PINN, V $\sin(x)$               & 10     & 0.1399 & 0.0955 & 0.2153 & 3330\\ 
PINN, Va                  & 4      & 0.5733 & 0.5321 & 2.3850 & 3406\\ 
PINN, Va                  & 6      & 0.0545 & 0.0399 & 0.1251 & 3485\\ 
PINN, Va                  & 8      & 0.0672 & 0.2315 & 0.063 & 3482\\ 
PINN, Va                 & 12     & 0.0833 & 0.0927 & 0.037 & 3759\\ 
\bottomrule
\end{tabular}
\end{table}

\subsection{3D cube case with forcing $\hat{s}=0.1$}

In the case of $\hat{s}=0.1$, the source function is more confined in the space. So far the lowest errors published where around 5\% compared to the finite element reference solution \citep{schoder2025physics}. After 200\,000 iterations, the network architectures presented in the previous section obtained errors in a range between 0.24\% to 1.6\%, when the network is adapting sufficiently to the wave field solution (see table \ref{tab:PINN_s01}). The standard PINN with higher scaling $\sin(4x)$, and $\sin(8x)$ exhibits larger relative errors. When investigating the number of training points, we detect an increase relative error below six points per wavelength.


\begin{table}[ht]
\centering
\caption{Summary of convergence error $e_\mathrm{rel,FEM}$ and training time $t$ for different complex 3D cube case with forcing $\hat{s}=1$ after 200\,000 iterations. }
\label{tab:PINN_s01}
\begin{tabular}{lccccc}
\toprule
\textbf{Name} & \textbf{$N_\mathrm{ppw}$} & \multicolumn{3}{c}{\textbf{$e_\mathrm{rel,FEM}$ (\%)}} & $t$ (s) \\
\midrule
PINN, $\sin(x)$                 & 10     & 0.38 & 0.38 & 0.24 & 4282\\ 
PINN, $\sin(2x)$                & 10     & 1.34 & 0.82 & 0.93 & 4485\\ 
PINN, $\sin(4x)$                & 10     & 2.79 & 4.61 & 3.68 & 4500\\ 
PINN, $\sin(8x)$                & 10     & 80.00 & 72.24 & 4.19 & 4488\\ 
PINN, Va                        & 10     & 0.85 & 1.21 & 1.57 & 3433\\ 
PINN, Vb                        & 10     & 0.92 & 0.91 & 0.75 & 3457\\ 
PINN, V $\sin(x)$               & 10     & 0.32 & 0.43 & 1.39 & 3318\\ 
PINN, Va                  & 4      & 77.19 & 75.30 & 95.00 & 3478\\ 
PINN, Va                  & 6      & 0.90 & 2.21 & 1.73 & 3432 \\ 
PINN, Va                  & 8      & 1.02 & 0.74 & 1.00 & 3480\\ 
PINN, Va                 & 12     & 0.41 & 0.95 & 0.78 & 3703\\ 
\bottomrule
\end{tabular}
\end{table}

\subsection{3D cube case with forcing $\hat{s}=0.01$}

\begin{figure}[htbp]
    \centering
    \begin{subfigure}[b]{0.3\textwidth}
        \includegraphics[width=\textwidth, trim=3cm 1cm 2cm 2cm, clip]{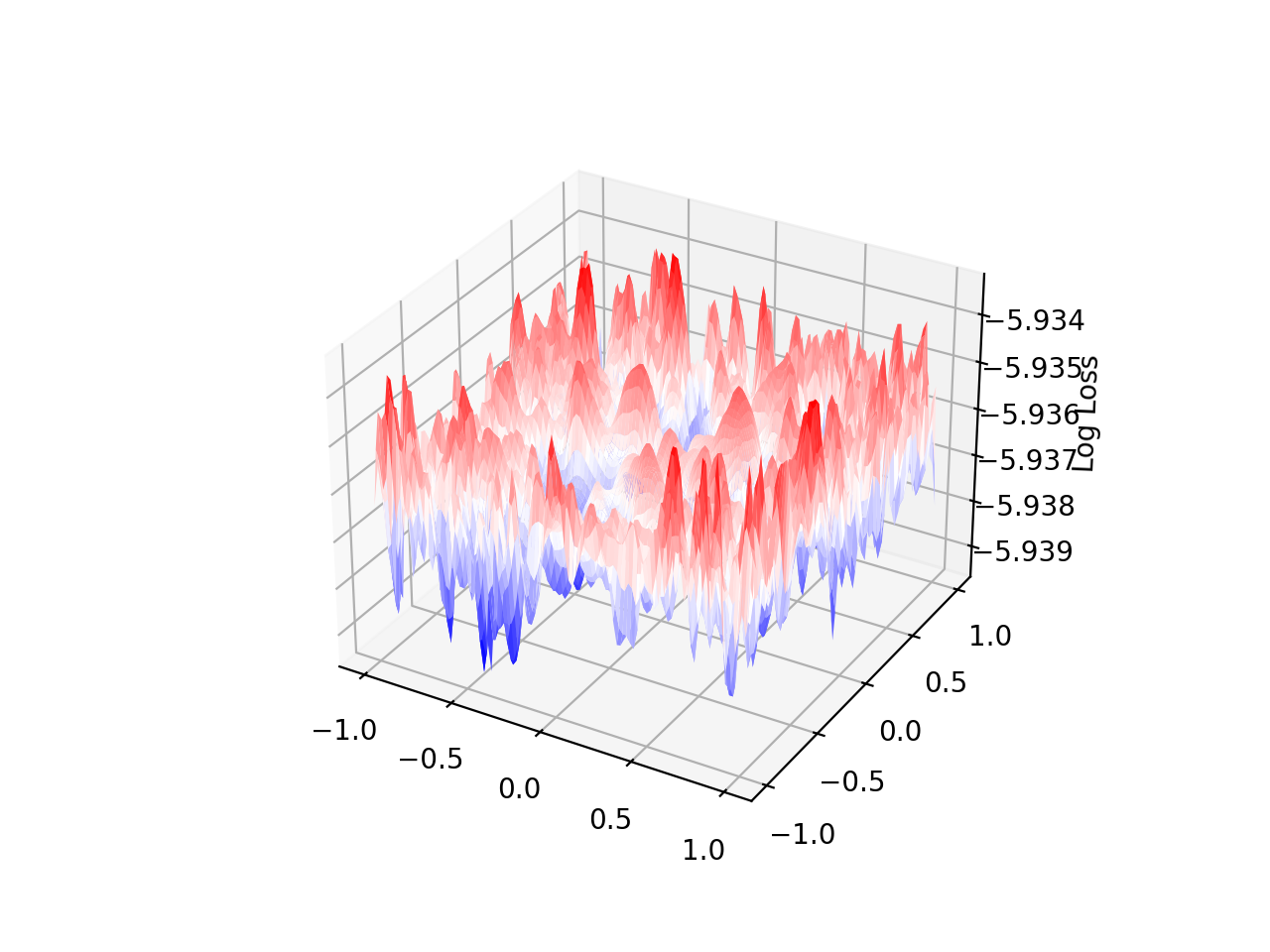}
        \caption{}
    \end{subfigure}
    \hfill
    \begin{subfigure}[b]{0.3\textwidth}
        \includegraphics[width=\textwidth, trim=3cm 1cm 2cm 2cm, clip]{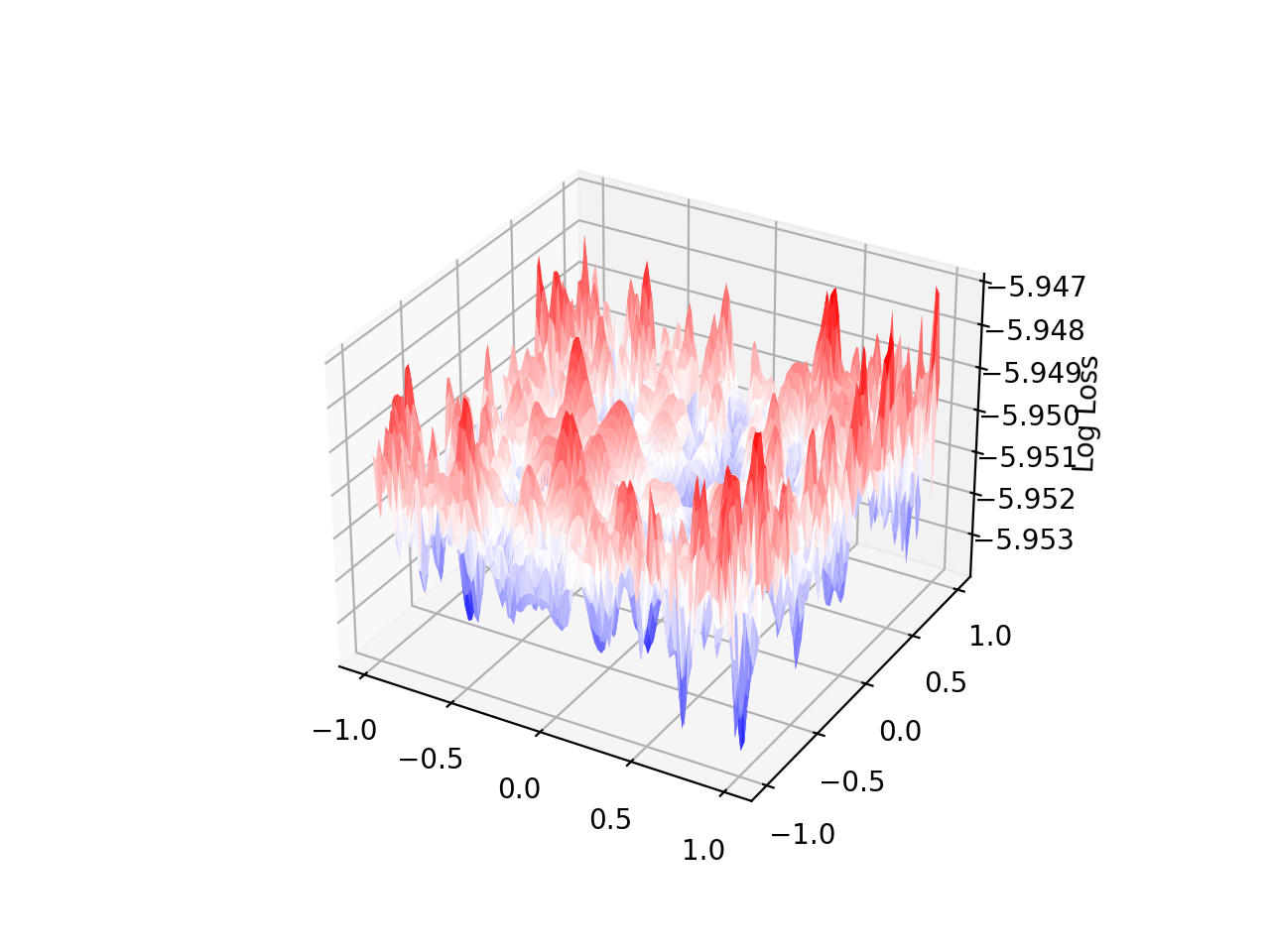}
        \caption{}
    \end{subfigure}
    \hfill
    \begin{subfigure}[b]{0.3\textwidth}
        \includegraphics[width=\textwidth, trim=3cm 1cm 2cm 2cm, clip]{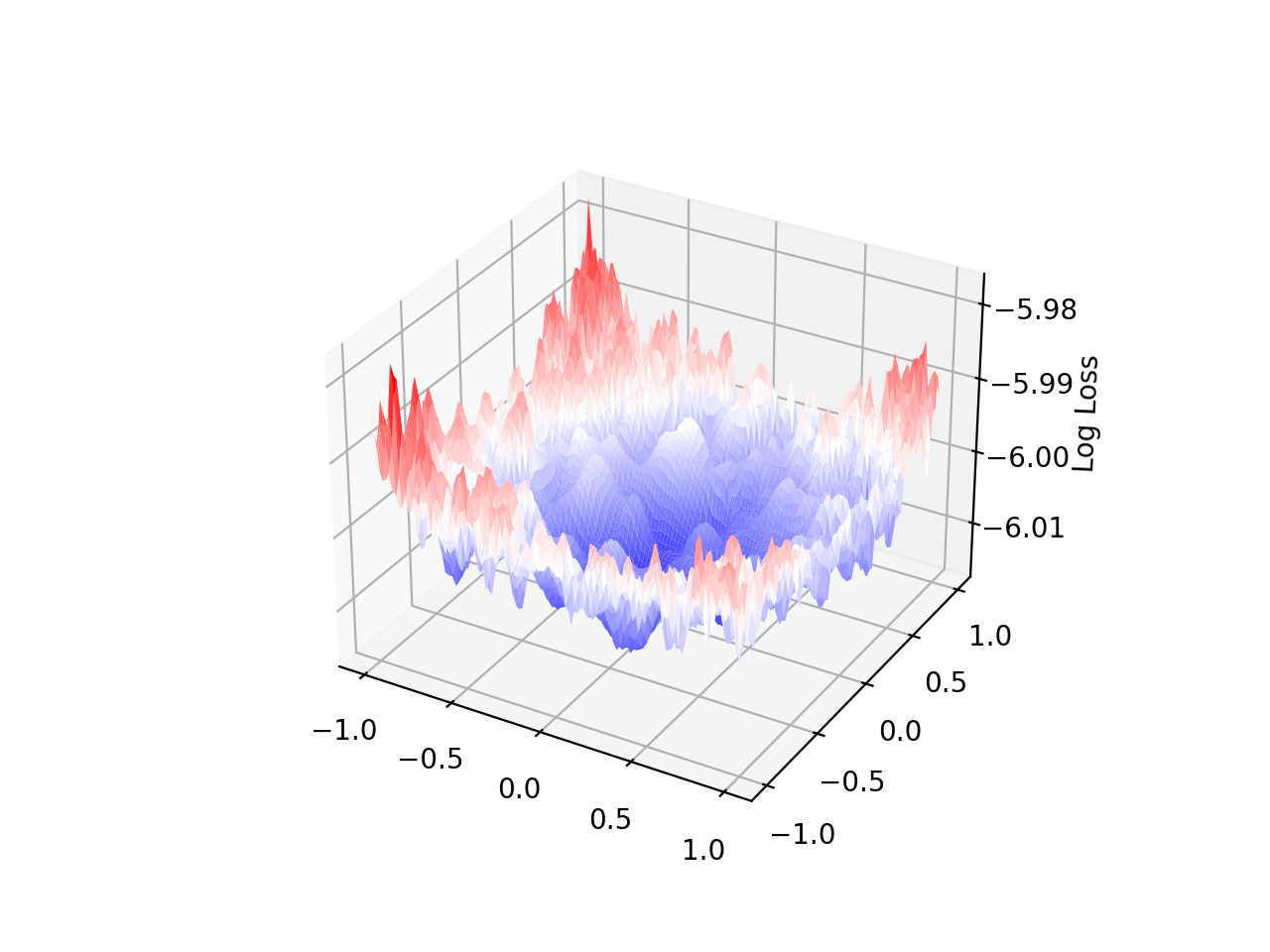}
        \caption{}
    \end{subfigure}
    \caption{Loss landscape for discrepancy learning, $\hat{s}=0.01$, no layers frozen:
    a) 8 b) 10 c) 12 points per wavelength. 
    } \label{fig:lossL_3d_s001}
\end{figure}

In the case $\hat{s}=0.01$, no convergence was obtained for the selected network types and the number of training points selected. This can be seen by the loss landscapes of the respective neural networks as a function of the points per wavelength (see figure \ref{fig:lossL_3d_s001}). The loss landscape oscillates very much for 8 and 10 points per wavelength, hinting that training will be difficult or impossible. At 12 points per wavelength, a global minimum in the loss landscape can be detected. Nevertheless, the loss landscape is very oscillatory. According to \cite{schoder2025physics}, at $\hat{s}=0.01$ the spatial frequency of interest is not the wavelength determined by the acoustic wave, instead one has to use the source variability defined by $\pi\hat{s} = 0.0314$m. This length scale is only resolved by one point per wavelength in the current setup. It may be interesting to investigate if adaptive refinement resolves the convergence issues here.

\subsection{Relative error when using discrepancy learning}

For discrepancy learning, the network has learned some improved version of the Green's function solution if it can adapt to the reflecting boundary condition (thus lower the error) compared to the reference finite element simulation of the acoustic wave field inside a room (see condition \eqref{eq:cond}). This was achieved for all setups, since the finally obtained error were well below the error of around 18\% ($\hat{s}=1$, see table \ref{table:DL_error_values_s1}) and 80\% ($\hat{s}=0.1$, see table \ref{table:DL_error_values_s01}). Since the reflection of the boundary condition might interact with the wave fields low and high spatial frequency content that will be learned by the neural network, we studied three training configurations. First, we allowed the pre-trained network that all its trainable parameters are adapted after activating the respective PINN loss function. Secondly, only the trainable parameters of the last two layers are used, as it is frequently applied in the concepts of transfer learning. Thirdly, only the first two layers parameters, which may be responsible for the global structures according to \cite{markidis2021old}. 

\begin{figure}[htbp]
    \centering
    \begin{subfigure}[b]{0.3\textwidth}
        \includegraphics[width=\textwidth, trim=3cm 1cm 2cm 2cm, clip]{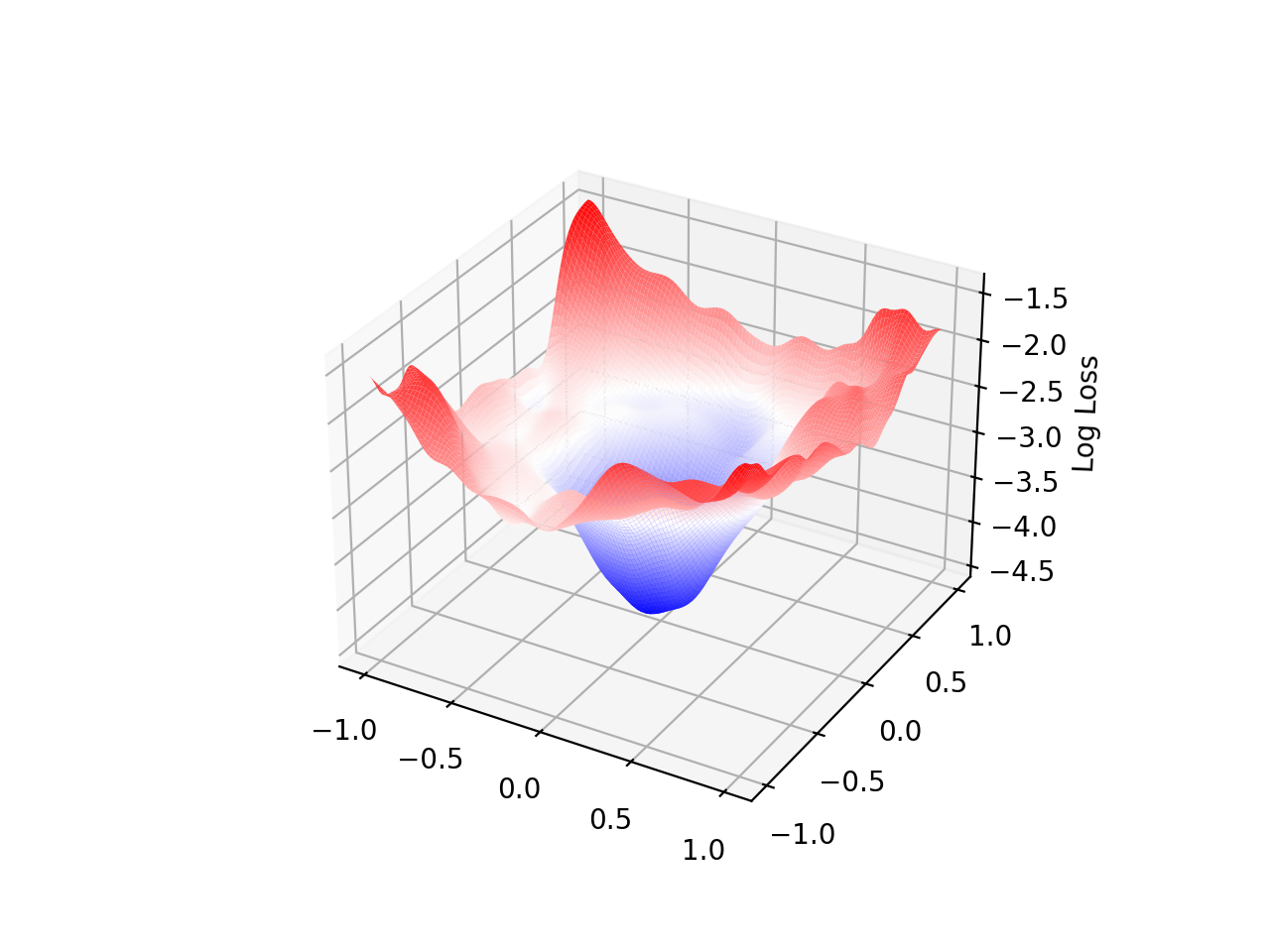}
        \caption{}
    \end{subfigure}
    \hfill
    \begin{subfigure}[b]{0.3\textwidth}
        \includegraphics[width=\textwidth, trim=3cm 1cm 2cm 2cm, clip]{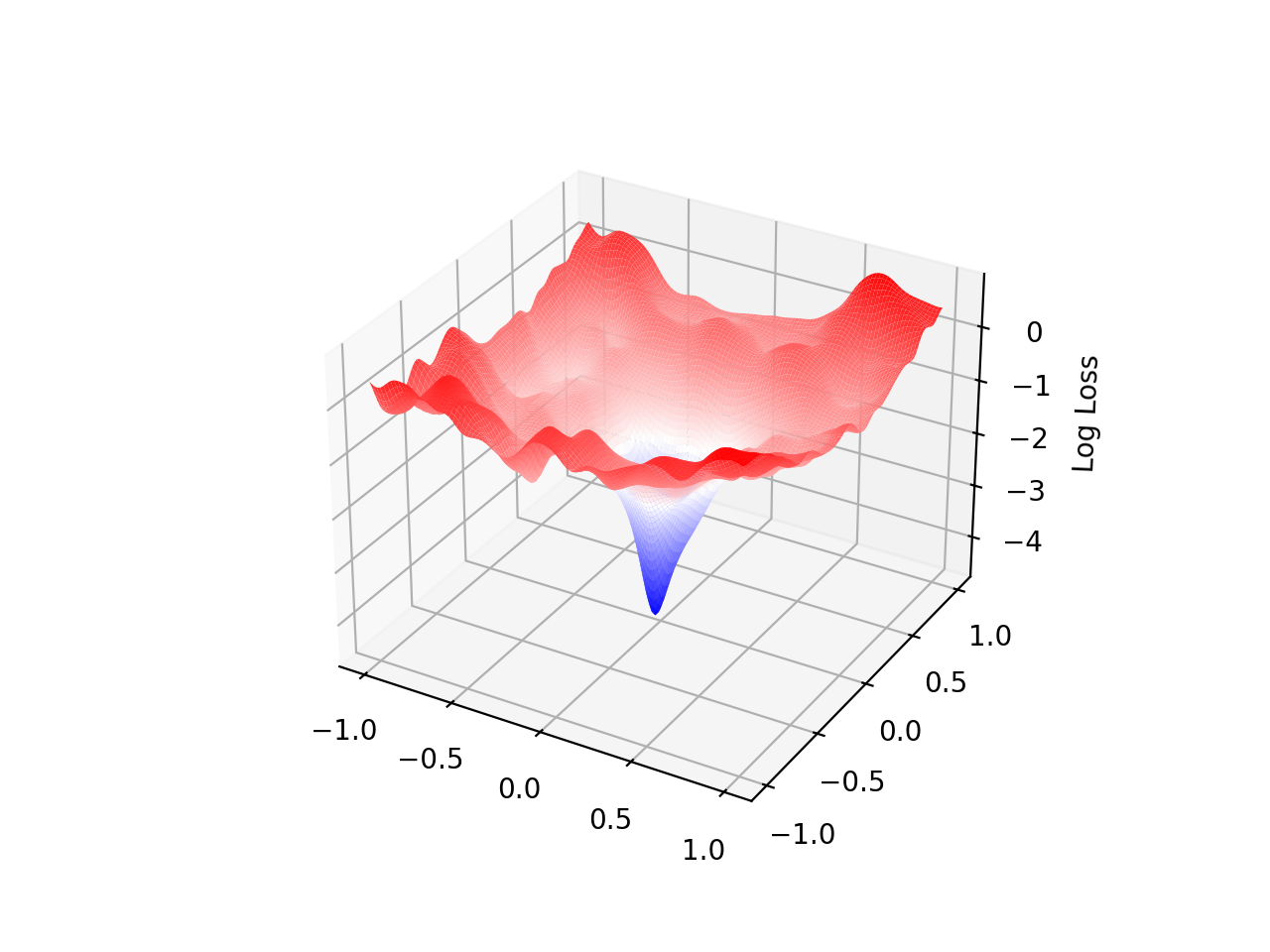}
        \caption{}
    \end{subfigure}
    \hfill
    \begin{subfigure}[b]{0.3\textwidth}
        \includegraphics[width=\textwidth, trim=3cm 1cm 2cm 2cm, clip]{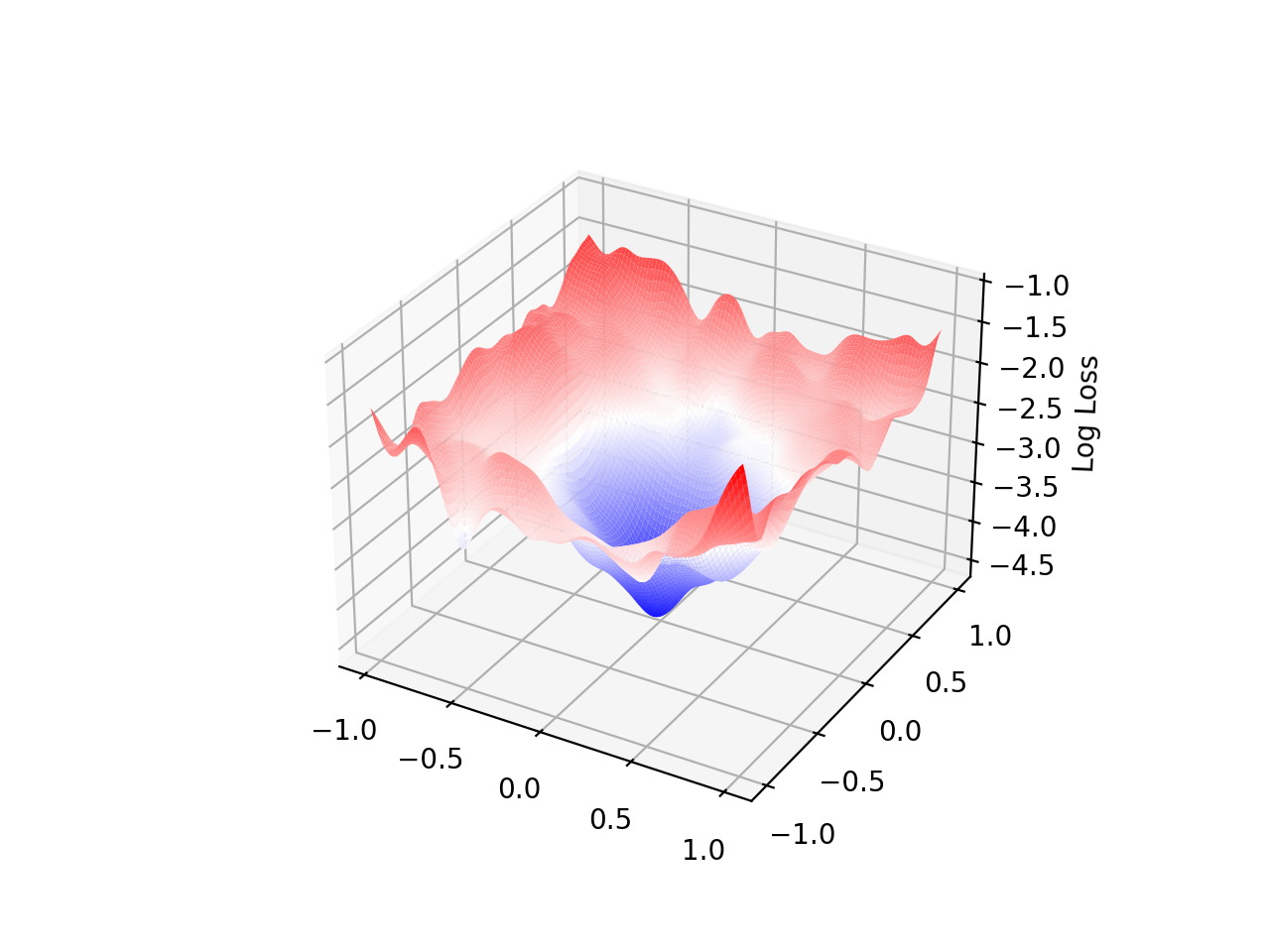}
        \caption{}
    \end{subfigure}
    \caption{Loss landscape for discrepancy learning for $\hat{s}=1$,
    where a) all b) first two c) last two layers were learnt by PINN
    and the other frozen.    
    }
    \label{fig:LLs1}
\end{figure}

In figure \ref{fig:LLs1}, we see that the loss landscapes of all layers and the last two layers learnt are similar, whereas the loss landscape for the first two layers learnt has a sharp transition from a flat exterior to a pronounced minimum. The pre-training was very beneficial in the time to convergence and reduced the training time to comparable accuracy by one order of magnitude. This was mainly because the Green's function data already fulfills the PDE-loss and the BC loss is the part that will be updated. Another observed effect during discrepancy learning was that optimization runs for the same pre-trained weights might end up in the same optimum (indicated by an x in the respective tables).

\subsection{3D rectangular cube case with forcing $\hat{s}=0.1$}

The final application example considers a typical non-uniformly extended cube representing a more realistic room of dimensions $\Omega = [0,1.3]$x$[0,1.0]$x$[0,0.7]$m$^3$ and $\hat{s}=0.1$. The convergence of the relative error (see table \ref{tab:PINN_s01_cube}) is comparable to the ones obtained for the same uniformly extended room and the same source shape parameter $\hat{s}=0.1$, presented in table \ref{tab:PINN_s01}. When using at least six points per wavelength the solution converges nicely.

\begin{table}[ht]
\centering
\caption{Summary of convergence error $e_\mathrm{rel,FEM}$ and training time $t$ for different complex 3D non-uniformly extended cube case with forcing $\hat{s}=0.1$ after 200\,000 iterations. }
\label{tab:PINN_s01_cube}
\begin{tabular}{lccccc}
\toprule
\textbf{Name} & \textbf{$N_\mathrm{ppw}$} & \multicolumn{3}{c}{\textbf{$e_\mathrm{rel,FEM}$ (\%)}}  & $t$ (s)\\
\midrule
PINN, $\sin(x)$                 & 10     & 0.72 & 0.523 & 0.807 & 4305\\ 
PINN, $\sin(2x)$                & 10     & 1.98 & 2.194 & 2.68 & 4473\\ 
PINN, $\sin(4x)$                & 10     & 6.60 & 6.87 & 8.63 & 4478\\ 
PINN, $\sin(8x)$                & 10     & 75.81 & 91.01 & 76.81 & 4478\\ 
PINN, Va                        & 10     & 1.69 & 1.59 & 1.86 & 3496\\ 
PINN, Vb                        & 10     & 1.05 & 1.70 & 1.11 & 3499\\ 
PINN, V $\sin(x)$               & 10     & 0.83 & 1.06 & 1.76 & 3272\\ 
PINN, Va                  & 4      & 97.76 & 97.18 & 96.44 & 3490\\ 
PINN, Va                  & 6      & 1.20 & 2.25 & 1.78 & 3462\\ 
PINN, Va                  & 8      & 3.04 & 2.18 & 1.50 & 3540\\ 
PINN, Va                 & 12     & 1.08 & 1.51 & 1.70 & 3726\\ 
\bottomrule
\end{tabular}
\end{table}

\section{Conclusions}
We assessed PINN architectures' convergence behavior for two-dimensional and three-dimensional acoustical problems 
and showed the limits of PINNs regarding realistic acoustic sources. 
Various source terms and two different room configurations were tested, using the relative $L^2$ error to a reference finite element simulation.

We would like to emphasize that room acoustics is an excellent discipline for this type of study: the numerical calculations that could be replaced by PINNs in the future are challenging due to the size of the systems of discretized equations. On the other hand, mathematically it is a well-studied problem and PINN loss does not risk the appearance of "parasitic" phenomena associated with some other types of PDEs (in particular the nonlinear ones, such as the Navier-Stokes equations).

In the course of investigation, we have confirmed that incorporating the knowledge of the underlying physics is beneficial for the whole training process. Specifically, we have discussed its involvement in adjusting the loss function (the argument of the sine -- see Sec \ref{sec:scaled}) and the pretraining on partly known solutions (see Sec \ref{sec:discrepancy}).
The quantitative results can be summarized as follows:
\begin{enumerate}
    \item Six training points per wavelength are necessary to obtain accurate results of the wave field.
    \item The convergence holds for a range of acoustic source fields calculated, as long as the second derivatives of the activation function remain nonzero. It is necessary for the underlying physics requirements. 
    \item By discrepancy learning of the Green's function first (pre-trained network), the time to solution from a forward problem can be reduced drastically.
\end{enumerate}

The limitations of our findings are based on the considered rectangular domain cases and the discussed source shape parameters. Whereas, some perspectives on the generalization were made during the investigations.


{
\small

\bibliographystyle{unsrt}
\bibliography{ref}
}

\newpage
\appendix

\section{Supplementary Material}
Additional Supplementary Material, including the source codes and the reference data is distributed as a zip file.
Additional Supplementary Material, including the source codes for training the models and obtaining the FEM reference data is distributed as a zip file.

\subsubsection{2D case: Supplementary Material on the Loss Landscape}
\begin{figure}[htbp]
    \centering
    \begin{subfigure}[b]{0.3\textwidth}
        \includegraphics[width=\textwidth, trim=3cm 1cm 2cm 2cm, clip]{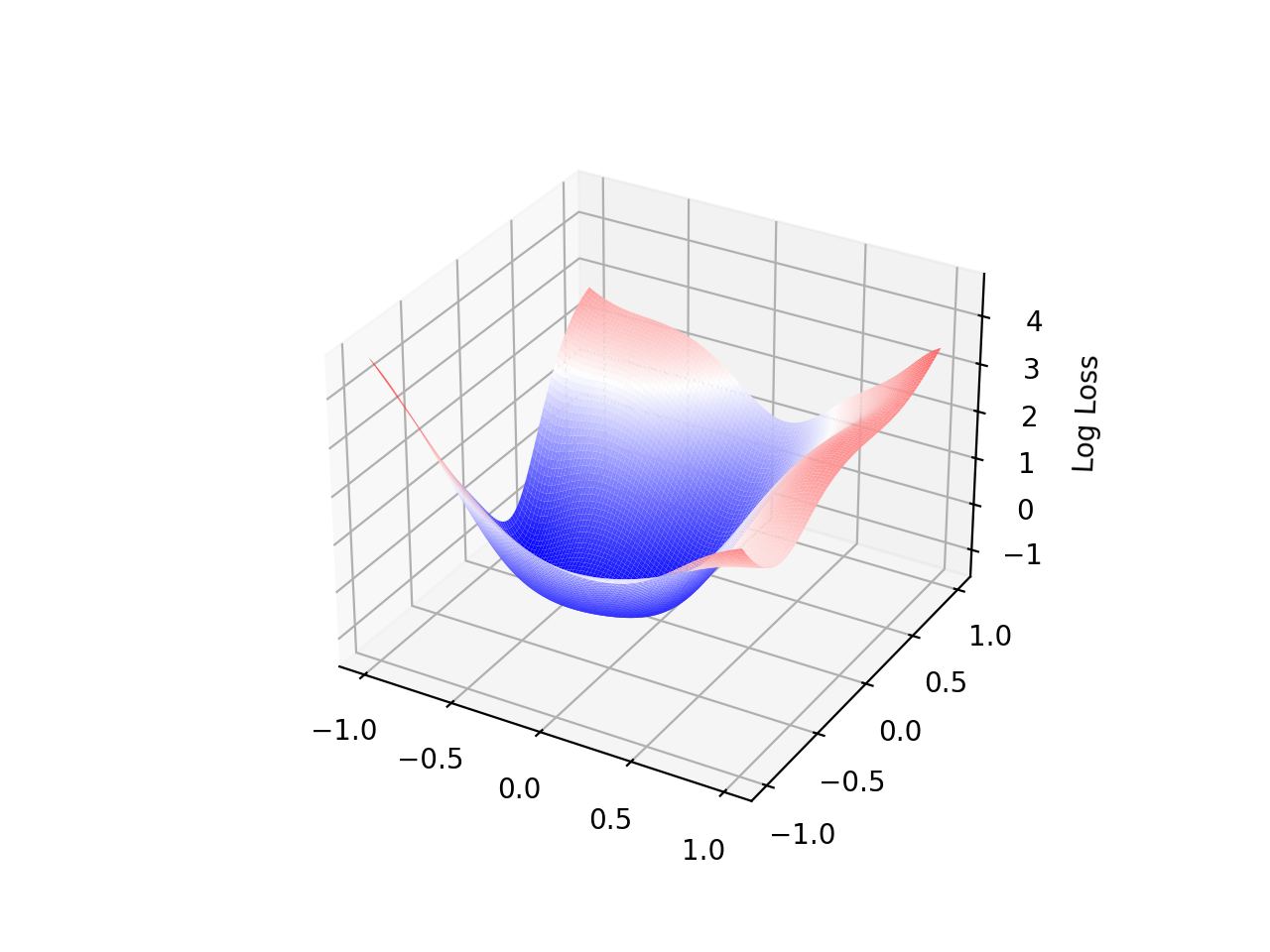}
        \caption{}
        \label{fig:plotA}
    \end{subfigure}
    \hfill
    \begin{subfigure}[b]{0.3\textwidth}
        \includegraphics[width=\textwidth, trim=3cm 1cm 2cm 2cm, clip]{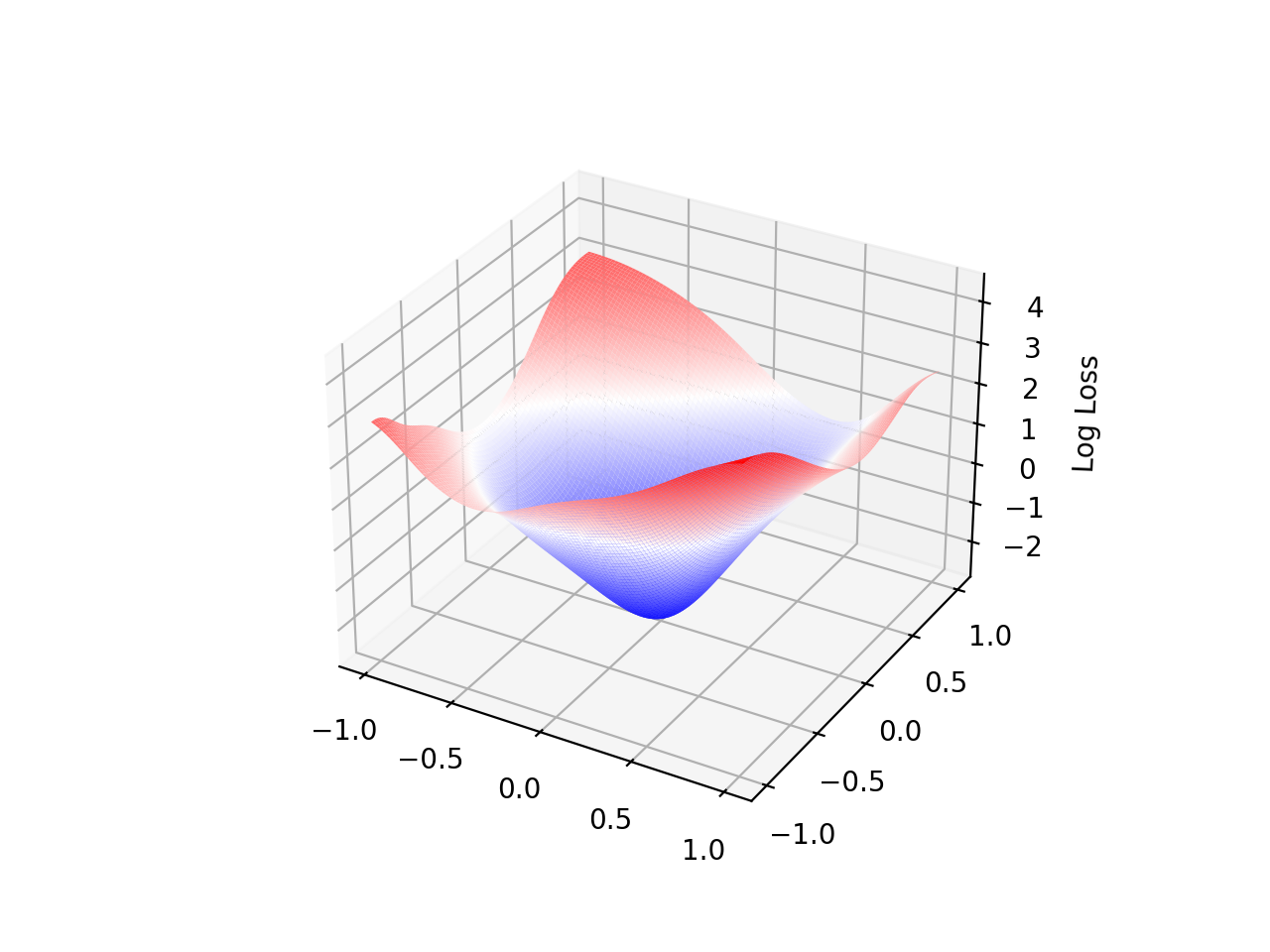}
        \caption{}
        \label{fig:plotB}
    \end{subfigure}
    \hfill
    \begin{subfigure}[b]{0.3\textwidth}
        \includegraphics[width=\textwidth, trim=3cm 1cm 2cm 2cm, clip]{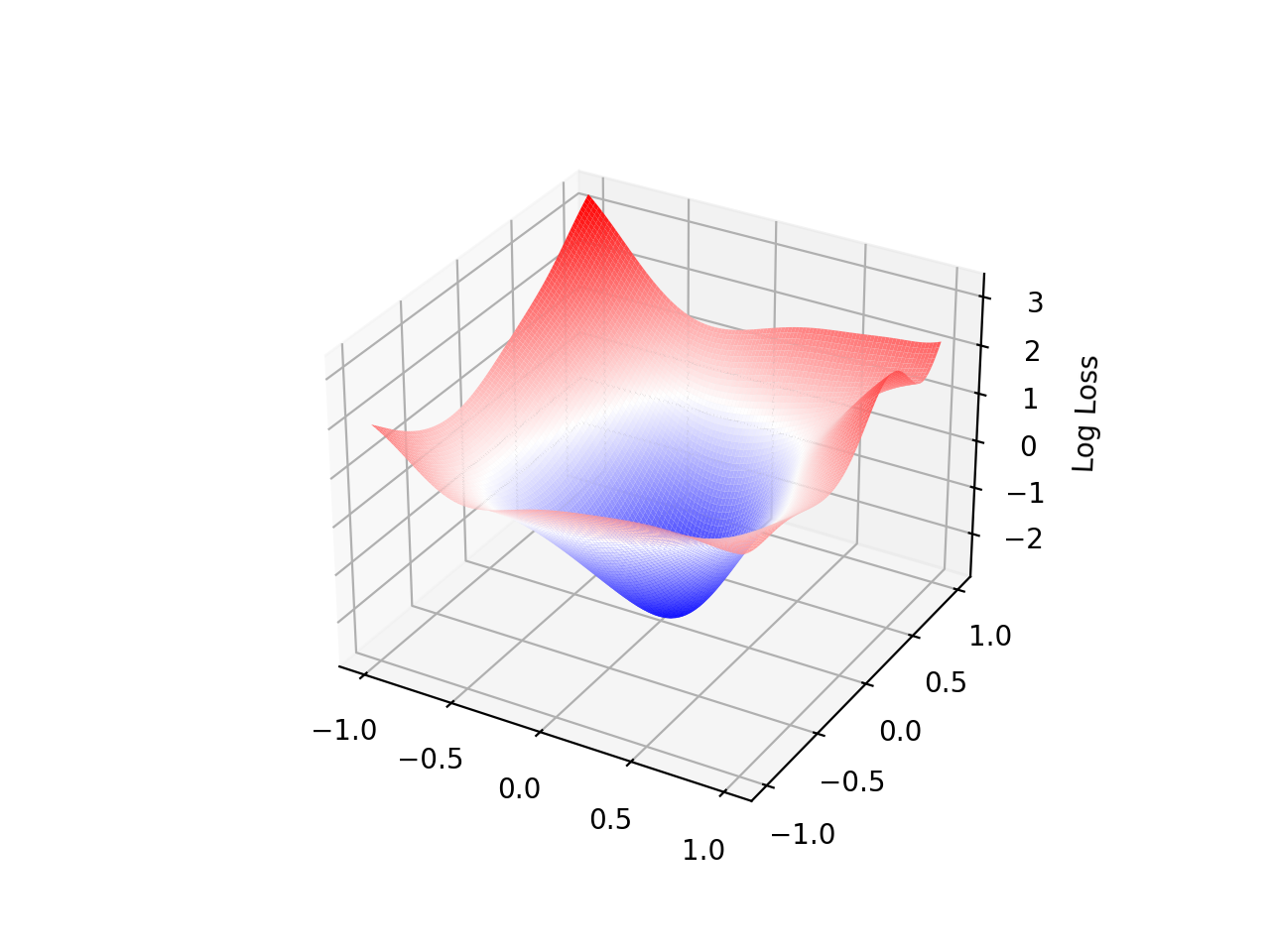}
        \caption{}
        \label{fig:plotC}
    \end{subfigure}
    \caption{Loss landscape of the 2D case, a) real valued speed of sound case modeled by a PINN with one boundary condition set, b) complex valued speed of sound case modeled by a PINN with one boundary condition set, and c) complex valued speed of sound case modeled by a PINN with four boundary condition set each for the real and imaginary part respectively. }
    \label{fig:lossL2d}
\end{figure}

\subsubsection{3D case: Supplementary Material on the Loss Landscape}

\begin{figure}[htbp]
    \centering
    \begin{subfigure}[b]{0.24\textwidth}
        \includegraphics[width=\textwidth, trim=3cm 1cm 2cm 2cm, clip]{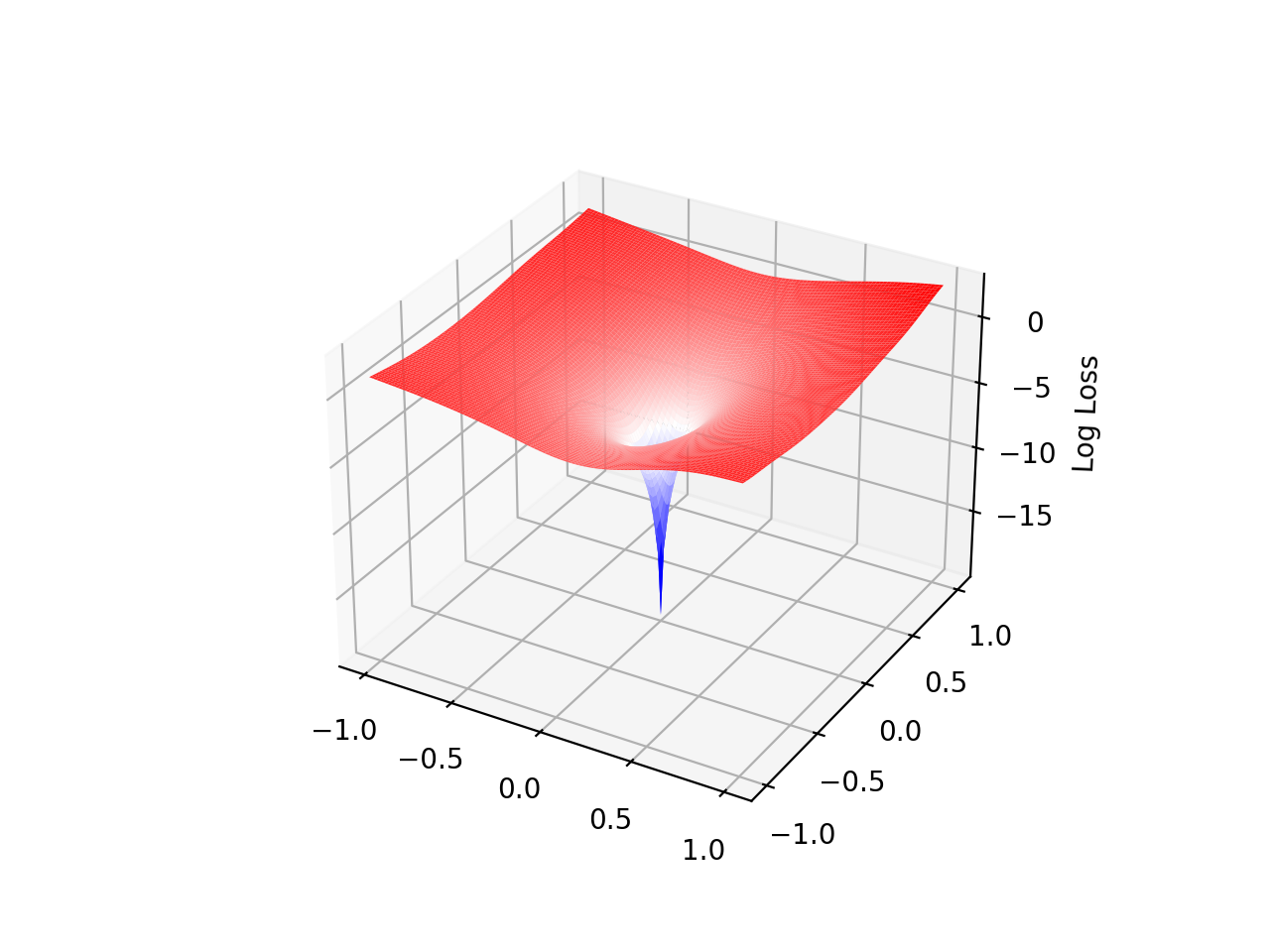}
        \caption{$\sin(x)$ }
    \end{subfigure}
    \hfill
    \begin{subfigure}[b]{0.24\textwidth}
        \includegraphics[width=\textwidth, trim=3cm 1cm 2cm 2cm, clip]{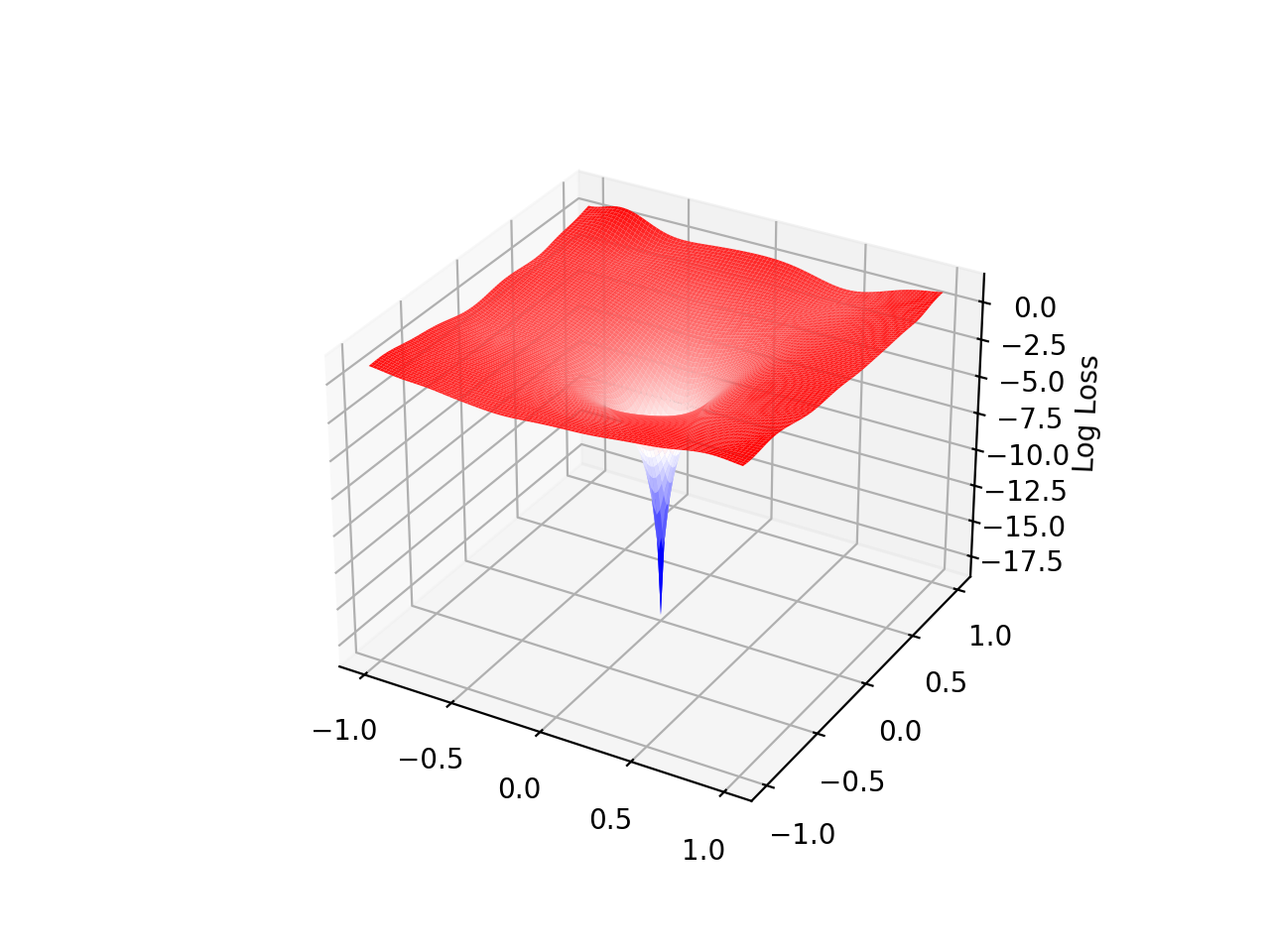}
        \caption{$\sin(2x)$}
    \end{subfigure}
    \hfill
    \begin{subfigure}[b]{0.24\textwidth}
        \includegraphics[width=\textwidth, trim=3cm 1cm 2cm 2cm, clip]{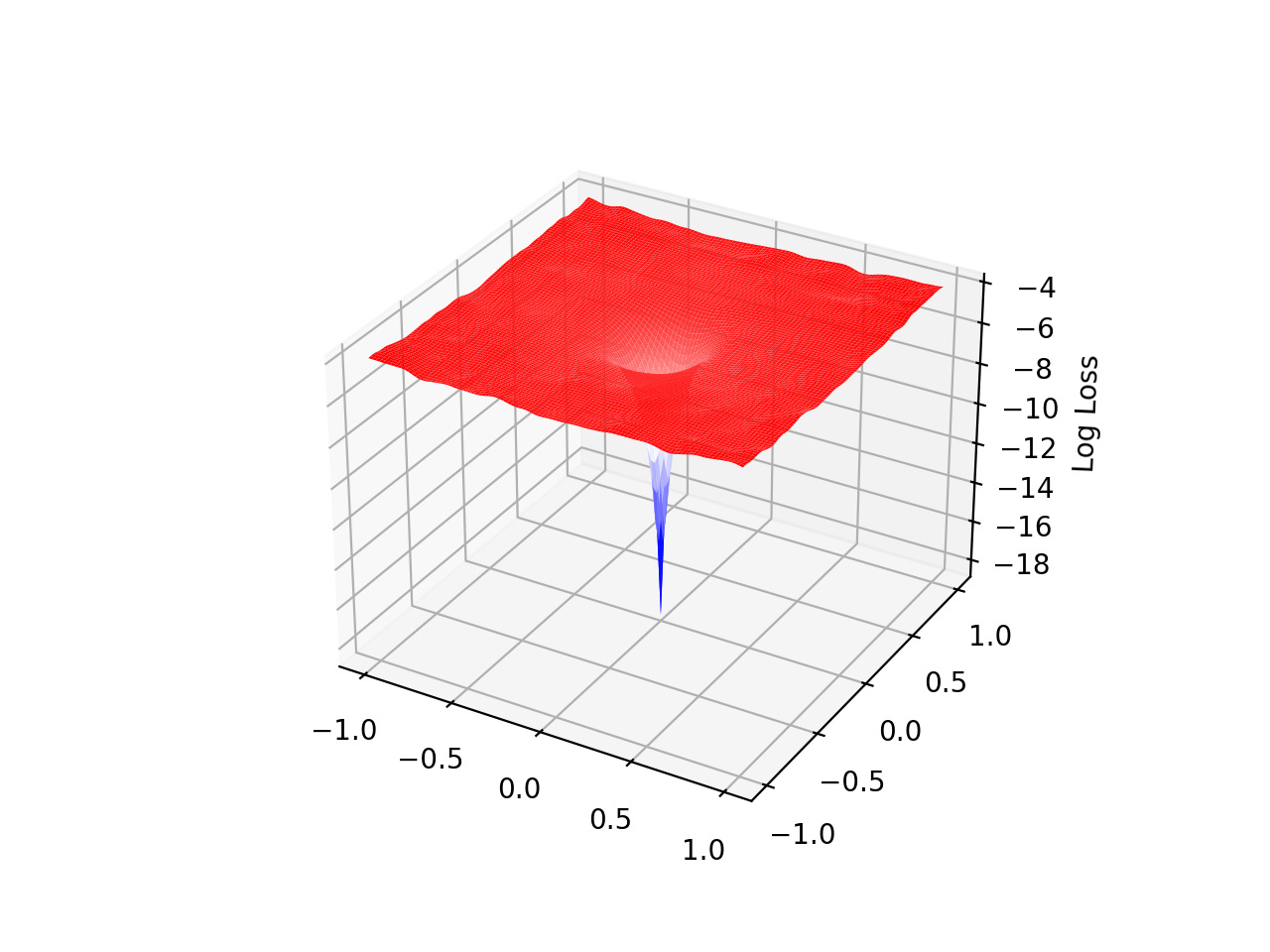}
        \caption{$\sin(4x)$}
    \end{subfigure}
        \hfill
    \begin{subfigure}[b]{0.24\textwidth}
        \includegraphics[width=\textwidth, trim=3cm 1cm 2cm 2cm, clip]{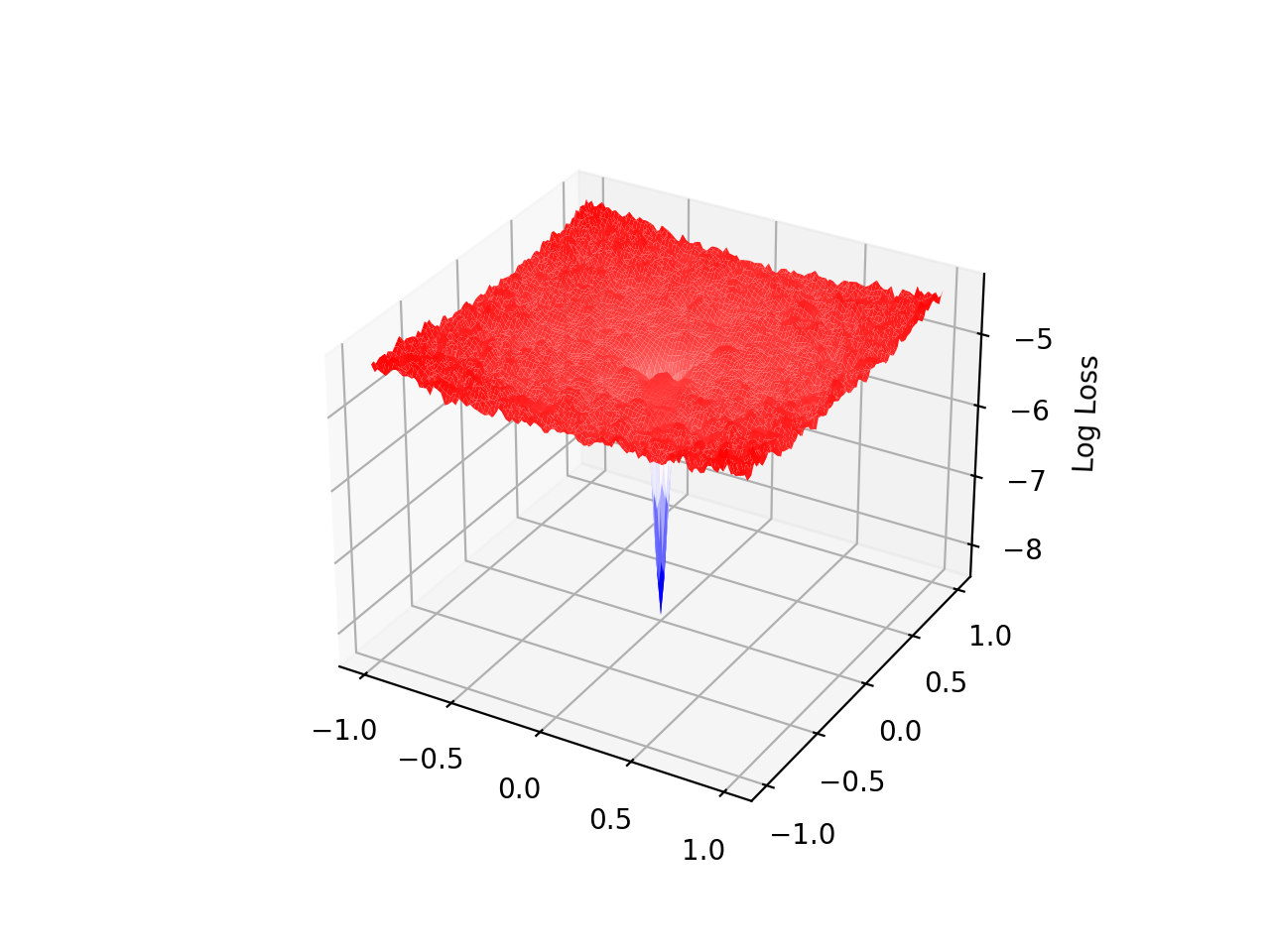}
        \caption{$\sin(8x)$}
    \end{subfigure}
    \caption{Loss landscape for purely data-trained network with $\hat{s}\rightarrow \infty$ and different activations functions.
    }\label{fig:lossL3dNN}
\end{figure}

Figure \ref{fig:lossL3dNN} shows the different loss landscapes. We see that for $\sin(8x)$ activation function it is highly oscillatory and therefore the optimizer has a hard time to find the global optimum and converge sufficiently.

\newpage
\subsection{Technical appendix} \label{sec:tech_appendix}

\subsubsection{Computational environment}
In this first section, we state the specifications of the computational environment such that the results are reproducible.



\begin{table}[h!]
\centering
\caption{Computational Environment PINN}
\begin{tabular}{|l|l|}
\hline
\textbf{Component}              & \textbf{Specification} \\
\hline
Programming language            & Python 3.11.11 \\
\hline
Deep learning library           & PyTorch 2.5.1 + CUDA 12.4 \\
\hline
PINN framework                  & DeepXDE 1.13.2 \\
\hline
Additional libraries            & NumPy 1.26.4, pyCFS 0.1.5 \\
\hline
CUDA version                    & 12.3 \\
\hline
GPU                             & NVIDIA A100-PCIE-40GB \\
\hline
GPU memory                      & 40 GB \\
\hline
GPU driver                      & 545.23.08 \\
\hline
Operating system                & Linux 4.18.0-477.10.1.el8\_8.x86\_64 \\
\hline
\end{tabular}
\label{tab:comp_env}
\end{table}

\begin{table}[h!]
\centering
\caption{Computational Environment FEM reference solution}
\begin{tabular}{|l|l|}
\hline
\textbf{Component}              & \textbf{Specification} \\
\hline
openCFS            & 24.03 \\
\hline
CPU                             & AMD Ryzen Pro 7 \\
\hline
Operating system                & Debian 12 \\
\hline
\end{tabular}
\end{table}

\subsubsection{2D case: Data point definition, hyperparameters, optimizer}\label{sec:TA:2dsetup}

\begin{table}[h!]
\centering
\caption{2D PINN Implementation}
\begin{tabular}{|l|l|}
\hline
\textbf{Hyperparameter}           & \textbf{Value / Description} \\
\hline
Optimizer                         & Adam \\
\hline
Learning rate                     & $1 \times 10^{-3}$ \\
\hline
Number of training iterations     & 70\,000 \\
\hline
Loss weights [BC; PDE]      & $[0.01; 1]$ (real), $[0.01, 0.0002; 1, 0.02]$ (real, imag)  \\
\hline
Network architecture              & FNN with 3 hidden layers, 150 neurons each \\
\hline
Activation function               & \texttt{sin} \\
\hline
Weight initializer                & Glorot uniform \\
\hline
Output transform                  & - \\
\hline
Training  points                & Randomly distributed, $n = \frac{n_{\mathrm{train}}}{\lambda}$ \\
\hline
Training  points domain         & $n^2$  \\
\hline
Training  points boundary       & $4 n$ \\
\hline
Number of test points ($L^2$ error)            & 40 $\times$ 40 \\
\hline
\end{tabular}
\label{tab:pinn_hyperparams_2d}
\end{table}

\newpage 
\subsubsection{3D case $\hat{s}\rightarrow \infty$: Data point definition, hyperparameters, optimizer}\label{sec:TA:3dsetup}
All definitions not given here, are the same as for the setup given in section \ref{sec:TA:2dsetup}. Table \ref{tab:pinn_hyperparams_3d_base} and table \ref{tab:NN_hyperparams_3d} are the network definitions regarding the errors provided in table \ref{tab:T1}.

\begin{table}[h!]
\centering
\caption{3D PINN Implementation - PINN 1BC/6BC  $\sin(x)$/ $\sin(2x)$/ $\sin(4x)$/ $\sin(8x)$ (base)}
\begin{tabular}{|l|l|}
\hline
\textbf{Hyperparameter}           & \textbf{Value / Description} \\
\hline
Number of training iterations     & 200\,000 \\
\hline
Loss weights [BC; PDE]      & $[5/k_0^2, 1/k_0^2; 1, 0.2]$ (real, imag)  \\
\hline
Training  points domain         & $n^3$  \\
\hline
Training  points boundary       & $6 n^2$ \\
\hline
Number of test points ($L^2$ error)            & 101 $\times$ 101 \\
\hline
\end{tabular}
\label{tab:pinn_hyperparams_3d_base}
\end{table}

\begin{table}[h!]
\centering
\caption{3D Supervised NN Implementation - NN  $\sin(x)$/ $\sin(2x)$/ $\sin(4x)$/ $\sin(8x)$ (base)}
\begin{tabular}{|l|l|}
\hline
\textbf{Hyperparameter}           & \textbf{Value / Description} \\
\hline
Number of training iterations     & 50\,000 \\
\hline
Loss       & MSE  \\
\hline
Training point locations         & 0.7\% (randomly sampled from FEM nodes)  \\
\hline
Training points values       & Analytic solution \\
\hline
Test points             & 0.3\% (randomly sampled from FEM nodes) \\
\hline
Number of test points ($L^2$ error)             & 101 $\times$ 101 \\
\hline
\end{tabular}
\label{tab:NN_hyperparams_3d}
\end{table}

\begin{table}[h!]
\centering
\caption{Discrepancy learning: 3D PINN Implementation - PINN 1BC  $\sin(x)$ (base)}
\begin{tabular}{|l|l|}
\hline
\textbf{Hyperparameter}           & \textbf{Value / Description} \\
\hline
Number of pre-training iterations (NN)    & 50\,000 \\
\hline
Number of training iterations (PINN)    & 40\,000 \\
\hline
Loss weights [BC; PDE]      & $[5/k_0^2, 1/k_0^2; 1, 0.2]$ (real, imag)  \\
\hline
Activation function               & \texttt{sin} \\
\hline
Training  points domain         & $n^3$  \\
\hline
Training  points boundary       & $6 n^2$ \\
\hline
Number of test points             & 101 $\times$ 101 \\
\hline
\end{tabular}
\label{tab:pinn_discrepancy_hyperparams_3d_base}
\end{table}

\newpage
\subsubsection{3D case $\hat{s}\rightarrow \infty$: Error tables}
\begin{table}[h!]
\centering
\caption{Relative FEM error $e_\mathrm{rel,FEM}$ for PINN in the complex case for different points-per-wavelength and frequencies. The optimization was infeasible due to the RAM requirements for a frequency of $\nu=8$, more than 10 points-per-wavelength. Therefore, no numbers for higher frequencies are presented.} \label{tab:PINN_ppw_3d_inf}
\label{tab:PINN_discrepancy_ppw_3d_inf}
\begin{tabular}{lccccccc}
\toprule
\textbf{Experiment Name} & $N_\mathrm{ppw}$ & \textbf{$\nu$} & \multicolumn{3}{c}{\textbf{$e_\mathrm{rel,FEM}$ (\%)}} & $t$ (s) \\
\midrule
PINN $\sin(x)$   & 4  & 1 & 21.79 & 22.97 & 28.16 & 3490\\
PINN $\sin(x)$   & 6  & 1 & 0.059 & 0.12 & 0.33 & 3523\\
PINN $\sin(x)$   & 8  & 1 & 0.046 & 0.14 & 0.46 & 3526\\
PINN $\sin(x)$  & 10 & 1 & 0.0072 & 0.021 & 0.23 & 3476\\
PINN $\sin(x)$  & 12 & 1 & 0.062 & 0.10 & 0.52 & 3489\\
\midrule
PINN $\sin(2x)$   & 4  & 2 & 47.48 & 70.84 & 79.30 & 3611\\
PINN $\sin(2x)$   & 6  & 2 & 0.065 & 0.11 & 0.15 & 3490\\
PINN $\sin(2x)$   & 8  & 2 & 0.044 & 0.071 & 0.11 & 3730\\
PINN $\sin(2x)$  & 10 & 2 & 0.043 & 0.058 & 0.074 & 4481 \\
PINN $\sin(2x)$  & 12 & 2 & 0.042 & 0.073 & 0.081 & 5989\\
\midrule
PINN $\sin(4x)$   & 4  & 2 & 52.90 & 55.47 & 55.72 & 3457\\
PINN $\sin(4x)$   & 6  & 2 & 12.51 & 15.60 & 16.06 & 3579\\
PINN $\sin(4x)$   & 8  & 2 & 0.034 & 0.074 & 0.16 & 3678\\
PINN $\sin(4x)$  & 10 & 2 & 0.027 & 0.039 & 0.053 & 4484\\
PINN $\sin(4x)$  & 12 & 2 & 0.029 & 0.033 & 0.038 & 6018\\
\midrule
PINN $\sin(4x)$   & 4  & 4 & 7.31 & 8.00 & 17.29 & 3675 \\
PINN $\sin(4x)$   & 6  & 4 & 0.62 & 0.63 & 0.90 & 5960\\
PINN $\sin(4x)$   & 8  & 4 & 0.42 & 0.52 & 0.58 & 12624 \\
PINN $\sin(4x)$  & 10 & 4 & 0.34 & 0.39 & 0.42 & 21753\\
PINN $\sin(4x)$  & 12 & 4 & 0.33 & 0.48 & 0.49 & 34333\\
\bottomrule
\end{tabular}
\end{table}

\begin{table}[h!]
\centering
\caption{Discrepancy learning: Relative FEM error $e_\mathrm{rel,FEM}$ for initial NN training and the subsequent training using PINN.}
\label{tab:PINN_discrepancy_ppw_3d_inf}
\begin{tabular}{lccccccc}
\toprule
\textbf{Experiment Name} & $N_\mathrm{ppw}$ & \textbf{$\nu$} & \multicolumn{3}{c}{\textbf{$e_\mathrm{rel,FEM}$ (\%)}}  & $t$ (s)\\
\midrule
NN $\sin(x)$   & 10  & 2 & 0.020 & 0.024 & 0.030 & 73\\
\midrule
PINN $\sin(x)$   & 4  & 2 & 1.75 & 0.59 & 2.35 & 612\\
PINN $\sin(x)$   & 6  & 2 & 0.12 & 0.094 & 0.11 & 599\\
PINN $\sin(x)$   & 8  & 2 & 0.065 & 0.056 & 0.082 & 639\\
PINN $\sin(x)$  & 10 & 2 & 0.023 & 0.057 & 0.071 & 779\\
PINN $\sin(x)$  & 12 & 2 & 0.023 & 0.10 & 0.055 & 1110\\
\midrule
Frozen, except first 2\\
PINN $\sin(x)$   & 4  & 2 & 1.85 & 1.23 & 9.70 & 603\\
PINN $\sin(x)$   & 6  & 2 & 0.85 & 0.39 & 4.17 & 607\\
PINN $\sin(x)$   & 8  & 2 & 1.17 & 0.50 & 4.09 & 604\\
PINN $\sin(x)$  & 10 & 2 & 0.73 & 0.39 & 4.16 & 767\\
PINN $\sin(x)$  & 12 & 2 & 0.59 & 0.45 & 8.52 & 1112\\
\midrule
Frozen, except last 2\\
PINN $\sin(x)$   & 4  & 2 & 0.79 & 0.50 & 2.84 & 574\\
PINN $\sin(x)$   & 6  & 2 & 0.18 & 0.11 & 0.085 & 573\\
PINN $\sin(x)$   & 8  & 2 & 0.090 & 0.099 & 0.12 & 606\\
PINN $\sin(x)$  & 10 & 2 & 0.062 & 0.090 & 0.12 & 779\\
PINN $\sin(x)$  & 12 & 2 & 0.062 & 0.090 & 0.11 & 802\\
\bottomrule
\end{tabular}
\end{table}

\newpage 
\subsubsection{3D case $\hat{s}=1$ discrepancy learning: Error tables}

\begin{table}[ht]
\centering
\caption{Error values and training times for 3D case $\hat{s}=1$ discrepancy learning}
\label{table:DL_error_values_s1}
\begin{tabular}{lccccc}
\hline
\textbf{Experiment Name} & $N_\mathrm{ppw}$ & \multicolumn{3}{c}{\textbf{$e_\mathrm{rel,FEM}$ (\%)}} & $t$ (s) \\
\midrule
NN $\sin(x)$   & 10  & 18.71 & 18.43 & 18.45 &138 \\
\midrule
PINN Va & 4 & 0.069 & 0.0477 & 0.069 & 617\\ 
PINN Va & 6  & 0.045 & 0.0141 & 0.0447 & 601\\ 
PINN Va & 8  & 0.0141 & 0.0467 & 0.037 & 623\\ 
PINN Va & 10 & 0.0467 & 0.0115 & 0.022 & 693\\ 
PINN Va & 12 & 0.0083 & 0.0083 & 0.0260 & 768 \\ 
\midrule
Frozen, except last 2\\
PINN Va & 4 & 0.1116 & 0.1320 & x & 593\\ 
PINN Va & 6 & 0.0338 & 0.026 & x & 591\\ 
PINN Va & 8 & 0.0142 & 0.011 & x & 597\\ 
PINN Va & 10 & 0.0106 & 0.016 & x & 606\\ 
PINN Va & 12 & 0.065 & x & x & 661\\ 
\hline
\end{tabular}
\end{table}

\subsubsection{3D case $\hat{s}=0.1$ discrepancy learning:: Error tables}

\begin{table}[ht]
\centering
\caption{Error values and training times for 3D case $\hat{s}=0.1$ discrepancy learning}
\label{table:DL_error_values_s01}
\begin{tabular}{lccccc}
\hline
\textbf{Experiment Name} & $N_\mathrm{ppw}$ & \multicolumn{3}{c}{\textbf{$e_\mathrm{rel,FEM}$ (\%)}} & $t$ (s) \\
\midrule
NN $\sin(x)$   & 10  & 79.62 & 79.43 & 79.60 & 142\\
\midrule
PINN Va & 4 & 1.58 & 15.63 & x & 613\\ 
PINN Va & 6  & 0.25 & 0.67 & x & 594\\ 
PINN Va & 8  & 0.55 & 0.55 & x & 608\\ 
PINN Va & 10 & 0.43 & 0.43 & x & 637\\ 
PINN Va & 12 & 0.32 & 0.32 & x & 650\\ 
\midrule
Frozen, except last 2\\
PINN Va & 4 & 8.53 & x & x & 620\\ 
PINN Va & 6 & 0.65 & 0.15 & 0.65 & 590\\ 
PINN Va & 8 & 0.31 & 0.133 & x & 596\\ 
PINN Va & 10 & 0.32 & 0.202 & 0.166 & 603\\ 
PINN Va & 12 & 0.33 & x & x & 605\\ 
\hline
\end{tabular}
\end{table}

\end{document}